\documentclass[fleqn,usenatbib]{mnras}

\usepackage[T1]{fontenc}

\DeclareRobustCommand{\VAN}[3]{#2}
\let\VANthebibliography\thebibliography
\def\thebibliography{\DeclareRobustCommand{\VAN}[3]{##3}\VANthebibliography}

\usepackage{graphicx}	
\usepackage{amsmath}	
\usepackage{amssymb}	

\usepackage{newtxtext,newtxmath}

\newcommand{\farc}{$^{\prime\prime}$}
\newcommand{\kalfa}{Fe $\rm K\alpha$}

\title[A physical X-ray model for the Circinus Galaxy]{A multiwavelength-motivated X-ray model for the Circinus Galaxy}

\author[C. Andonie et al.]{Carolina Andonie,$^{1,2,3}$
Claudio Ricci$^{4,5}$, 
St\'ephane Paltani$^{6}$, 
Patricia Ar\'evalo$^7$, 
\newauthor
Ezequiel Treister$^2$,
Franz Bauer$^{2,3,8}$,  
and Marko Stalevski$^{9,10}$\\
$^{1}$Centre for Extragalactic Astronomy, Department of Physics, Durham University, South Road, DH1 3LE, Durham, UK \\
$^2$Instituto de Astrof\'isica, Facultad de F\'isica, Pontificia Universidad Cat\'olica de Chile, Santiago , Chile \\
$^3$Millennium Institute of Astrophysics, Vicuña Mackenna 4860, Macul, Santiago, Chile\\
$^4$  N\'ucleo de Astronom\'ia de la Facultad de Ingenier\'ia, Universidad Diego Portales, Av. Ej\'ercito Libertador 441, Santiago, Chile\\
$^5$ Kavli Institute for Astronomy and Astrophysics, Peking University, Beijing 100871, China\\
$^6$ Department of Astronomy, University of Geneva, ch. d'Ecogia 16, CH-1290 Versoix, Switzerland \\
$^{7}$Instituto de F\'isica y Astronom\'ia, Facultad de Ciencias, Universidad de Valpara\'iso, Gran Bretana N1111, Playa Ancha, Valpara\'iso, Chile\\
$^{8}$Space Science Institute, 4750 Walnut Street, Suite 205, Boulder, Colorado 80301, USA\\
$^9$ Astronomical Observatory, Volgina 7, 11060 Belgrade, Serbia\\ 
$^{10}$ Sterrenkundig Observatorium, Universiteit Ghent, Krijgslaan 281--S9, Ghent, B-9000, Belgium\\
}

\date{Accepted XXX. Received YYY; in original form ZZZ}

\pubyear{2015}

\begin{document}
\label{firstpage}
\pagerange{\pageref{firstpage}--\pageref{lastpage}}
\maketitle

\begin{abstract}

Reprocessed X-ray emission in Active Galactic Nuclei (AGN) can provide fundamental information about the circumnuclear environments of supermassive black holes. Recent mid-infrared studies have shown evidence of an extended dusty structure perpendicular to the torus plane. In this work, we build a self-consistent X-ray model for the Circinus Galaxy including the different physical components observed at different wavelengths and needed to reproduce both the morphological and spectral properties of this object in the mid-infrared. The model consists of four components: the accretion disk, the broad line region (BLR), a flared disk in the equatorial plane and a hollow cone in the polar direction. Our final model reproduces well the 3--70 keV {\it Chandra} and {\it NuSTAR} spectra of Circinus, including the complex Fe K$\alpha$ zone and the spectral curvature, although several additional Gaussian lines, associated to either ionized iron or to broadened Fe K$\alpha$/K$\beta$ lines, are needed. We find that the flared disk is Compton thick ($ N_{\rm H,d}= \rm 1.01^{+0.03}_{-0.24}\times 10^{25}\: cm^{-2}$) and geometrically thick ($CF=0.55^{+0.01}_{-0.05}$), and that the hollow cone has a Compton-thin column density ($ N_{\rm H,c}= \rm 2.18^{+0.47}_{-0.43}\times 10^{23}\: cm^{-2}$), which is consistent with the values inferred by mid-infrared studies. Including also the BLR, the effective line of sight column density is $ N_{\rm H}= \rm 1.47^{+0.03}_{-0.24}\times 10^{25}\: cm^{-2}$. This approach to X-ray modelling, i.e. including all the different reprocessing structures, will be very important to fully exploit data from future X-ray missions.

\end{abstract}




\section{Introduction}

It is widely known that most massive galaxies host a supermassive black hole (SMBH) at their center \citep[e.g.][]{1998AJ....115.2285M}. During the phase in which these SMBHs accrete matter they are observed as active galactic nuclei (AGN). The central SMBH can be heavily obscured by large columns of gas and dust \citep{1991PASJ...43L..37A}. According to the simplest AGN unification models (e.g., \citealp{Antonucci:1993fu,Netzer:2015rev,Ramos-Almeida:2017qq}) the obscuring material is associated with a parsec-scale dusty torus in the same equatorial plane as the accretion disk and the broad line region (BLR). In AGN observed pole-on with respect to this torus, one would be able to directly observe the emission from the accretion disk and the BLR (type 1 or unobscured AGN), while in sources observed edge-on these components would not be visible (type 2 or obscured AGN). 

An important feature in AGN is their X-ray emission, which originates in a compact region, the corona, where hot electrons transfer their energy through inverse Compton scattering to optical/UV photons produced in the accretion disk (\citealp[e.g.,][]{1991ApJ...380L..51H}). In their X-ray spectra, AGN also show signatures of reprocessing by the circumnuclear material, which has often been associated to the torus \citep[e.g.,][]{1994MNRAS.267..743G,2005A&A...444..119G}. About 70$\%$ of all AGN in the local Universe are obscured \citep[e.g.,][]{2015ApJ...815L..13R,Ricci2017ss}. When the obscurer is Compton-thick ($N_{\rm H}\geq10^{24}\rm cm^{-2}$) it depletes most of the X-ray continuum \citep[e.g.][]{2000MNRAS.318..173M}, thereby providing a clearer view of the reprocessed X-ray radiation produced in the circumnuclear gas.  Therefore, heavily obscured AGN are potentially great laboratories to study the circumnuclear environment of the central SMBH in the X-ray band.

The dusty torus not only reprocesses the X-ray continuum, but it also absorbs the optical/UV emission produced in the accretion disc, re-emitting it in the infrared (IR). With the recent advances of IR interferometry, it is now possible to resolve the innermost regions of nearby AGN \citep[e.g.,][]{2004Natur.429...47J,2013A&A...558A.149B,2016A&A...591A..47L}.
Unexpectedly, over the past decade, an extended mid-infrared (MIR) component, elongated in the polar direction, has been detected in several sources (e.g., \citealp{2012ApJ...755..149H,2014A&A...563A..82T}). This elongated emission has been found both on parsec scale by high-resolution interferometric studies (\citealp[e.g.,][]{2012ApJ...755..149H,2013A&A...558A.149B,2016A&A...591A..47L,2019ApJ...886...55L}), and on scales of tens to hundreds of parsecs by single-dish observations (e.g., \citealp{2016ApJ...822..109A,2019MNRAS.489.2177A}). Recently, \citet{2016ApJ...822..109A} and \citet{2019MNRAS.489.2177A} used data from the upgraded
Very Large Telescope (VLT) Imager and Spectrometer for mid-InfraRed (VISIR; \citealp{2004Msngr.117...12L}) to study the extent of the polar dust, and found that 26 out of 30 obscured nearby AGN show polar MIR emission extending up to tens of parsec, which accounts for $83\%$ of the total MIR AGN emission. Radiation pressure on dusty gas \citep[e.g.,][]{2006MNRAS.373L..16F,2007MNRAS.380.1172H,2008MNRAS.385L..43F,2020Venanzi} has been shown to be very effective in cleaning up the close environments of AGN \citep{Ricci2017ss}, and could lead to the creation of a dusty polar component. \citet{2019ApJ...886...55L} studied a sample of 33 AGN with the Very Large Telescope Interferometer (VLTI)/MID-infrared Interferometer (MIDI), and found a tentative positive relation between fraction of dust in the polar component and the Eddington ratio, which suggests that indeed this elongated component might be driven by radiation pressure (see also \citealp{Ricci2017ss,2020Venanzi}). All this evidence indicates that the fiducial torus is not the only dusty structure acting as reprocessor, and a polar dust component is needed to explain the observed IR properties of AGN. Past studies have also found elongated emission in the X-rays \citep[e.g.,][]{2018Fabbiano,2018BFabbiano}, using observations from the \textit{Chandra} observatory \citep{2000Weisskopf}. This emission could be related, at least in some cases, to the polar dust observed in the MIR (e.g., \citealp{2019MNRAS.490.4344L}). It has been also suggested that the polar component might be responsible for some of the narrow emission features observed at $E\lesssim 3$\,keV in heavily obscured AGN, such as the Si K$\alpha$ line (e.g., \citealp{2019MNRAS.490.4344L,McKaig:2021vw}).

The AGN in which the polar component has been studied most in detail is the Circinus Galaxy (hereafter Circinus). The proximity of Circinus ($\sim 4.2\rm \: Mpc$ \citealp{1977A&A....55..445F}), has turned it into a great laboratory to study the AGN phenomenon \citep[e.g.,][]{1996MNRAS.281L..69M,2013Marinucci,2013ApJ...768L..38H,2014A&A...563A..82T}. Using high-quality MIR images, obtained with VISIR, \citeauthor{2017MNRAS.472.3854S} (\citeyear{2017MNRAS.472.3854S}; hereafter S17) were able to explain the tens-of-parsec dust-emitting regions in Circinus, as a compact flared-disk in the equatorial plane and a hollow cone-like region extending up to 40 pc in the polar direction. Following this, \citeauthor{2019MNRAS.484.3334S} (\citeyear{2019MNRAS.484.3334S}; hereafter S19) refined their results, using higher-resolution data from the VLTI/MIDI, explaining the extended emission in the central few pc as a hyperboloid shell. 
Using $\rm H_2O$ maser emission ($\lambda = 1.3\rm \:cm$) maps, \citet{2003Greenhill} found that Circinus has a warped accretion disk located at 0.1 pc from the center, and suggested that the whole system is also edge-on. This has been supported by several X-ray studies that have concluded that the AGN of Circinus is obscured by Compton-thick column densities and is observed edge-on \citep[e.g.,][]{2014ApJ7...91...81A,2019A&A...629A..16B,2021Uematsu}.

While detailed studies of the geometry of the reprocessing material in Circinus have been performed in the MIR, in the X-ray band a study that includes all the different AGN components has not been carried out yet. The aim of our work is to use the IR models proposed by \citet{2017MNRAS.472.3854S, 2019MNRAS.484.3334S} to explain the X-ray emission of Circinus. A few previous studies have combined IR and X-ray spectral analysis, performed by considering similar models in the two energy regimes, typically a smooth or a clumpy torus (e.g., 
\citealp{Farrah:2016nw,Lanz:2019ey,Esparza-Arredondo:2019ux,Esparza-Arredondo:2021ep}). Most of the current X-ray models of reprocessing material \citep[e.g][]{2009MNRAS.397.1549M,2011MNRAS.413.1206B,2014ApJ...787...52L,2018ApJ...854...42B,2019A&A...629A..16B,2019ApJ...877...95T,2020ApJ...897....2T}  only include the torus as the reprocessor. In order to perform an X-ray analysis of Circinus as accurate and self-consistent as possible we build a physical X-ray model based on the aforementioned MIR models. In our model we also include the accretion disk and the BLR, since they are expected to contribute to the scattered X-ray radiation. This is the first time an X-ray spectral model includes all the AGN components, and is built based on observations carried out at different energies.     

The paper is organised as follows. In \S\ref{sec:models} we illustrate the details of the MIR models proposed by \citet{2017MNRAS.472.3854S, 2019MNRAS.484.3334S} and our X-ray model. In \S\ref{sec:xray} and \S\ref{sec:xrayspectralfitting} we present the X-ray data and the spectral fitting procedure, respectively. In \S\ref{sect:discussion} we discuss our results, while in \S\ref{sec:conclusions} we present our conclusions.

\section{The X-ray model of Circinus} \label{sec:models}
\subsection{The infrared models}\label{sec:IRmodel}

S17 and S19 recently proposed two models to explain recent MIR observations of Circinus.
In S17 the authors explain the MIR emission on scales of tens of parsecs, comparing the MIR morphology and spectral energy distribution (SED) to expectations from a physical model. For the  morphological comparison they used high-fidelity VLT/VISIR images, while for the SED they included data from \textit{Spitzer}/IRS and VLT/MIDI (for further details see Table\,\,1 of S17). They concluded that the model which best reproduces the data is a flared disk in the equatorial plane and a hollow cone for the polar dust. The parameters of the disc are the angular half width (measured from the equatorial plane to the edge of the disc), the outer radius of the disc and the radial optical depth. The parameters of the cone are the radial extension, the half opening angle (measured from the polar axis of the system), the angular full width and the radial optical depth. The details of this model are given in Table\,\ref{t:ir_par1}, and are illustrated in Figure\,\,3 of S17. 
\begin{table}
\centering
 \begin{tabular}{l c c} 
\multicolumn{3}{c}{\bf \Large IR model} \\
 \noalign{\medskip}
 \hline
 \hline
 \noalign{\medskip}

\multicolumn{3}{c}{\bf \large Disc} \\
 \noalign{\smallskip}
& {\bf S17} &{\bf S19}\\
 \hline
 \noalign{\smallskip}
Angular half width & $20^{\circ}$ &  $5^{\circ}$ \\
 \noalign{\smallskip}
Outer radius   & 1.5 pc & 3 pc\\
 \noalign{\smallskip}
Radial optical depth ($\tau_{V}$)   & $47$ & $141$ \\
 \noalign{\bigskip}
\multicolumn{3}{c}{\bf \large Cone/hyperboloid} \\
 \noalign{\smallskip}
& {\bf S17} &{\bf S19}\\
 \hline
 \noalign{\smallskip}
Half opening angle & $40^{\circ}$ & $30^{\circ}$ \\
 \noalign{\smallskip}
Angular full width  & $10^{\circ}$ & $1^{\circ}$ \\
 \noalign{\smallskip}
Radial extension & 40 pc & -- \\
 \noalign{\smallskip}
Outer wall position ($a_{\rm out}$)  & --  & $0.6\rm\,pc$ \\
\noalign{\smallskip}
Radial optical depth ($\tau_{V}$)  & $1$ & $53.5$ \\
\hline
 \end{tabular}
 \caption{Parameters of the IR model at large and small scales. For more details, see \citet{2017MNRAS.472.3854S} and \citet{2019MNRAS.484.3334S}. Further details can be found in \S\ref{sec:IRmodel}. Note that (a) VISIR data did not have constraining power over the small scale structure (i.e. the disc parameters), (b) the angular width and opening angle of the hyperboloid shell are expected to increase with the distance and to match the values of the cone parameters at larger scale, and (c) the radial optical depth of the two models are not directly comparable (see the discussion of the caveats in the section 4.1.4 of S19).} \label{t:ir_par1}
\end{table}

In S19, the results for the parsec-scale geometry of the circumnuclear dust are refined, thanks to the analysis of high-resolution interferometric VLTI/MIDI data. The authors compare the correlated flux for different baselines, phases and wavelength (see Section\,4.1 of S19), finding that the model which best reproduces the MIR emission in the central few parsecs is a a geometrically thin flared disk for the equatorial component, and a hyperboloid shell for the polar component. The parameters of the disc are the same as in S17, while those of the hyperboloid are the half opening angle, the angular full width, the outer wall position\footnote{distance between the center and the outer wall of the hyperboloid, in the equatorial plane} and the radial optical depth. To account for the near-IR flux, the hyperboloid shell in the final S19 model was made to be clumpy. The details of the model are summarized in Table\,\ref{t:ir_par1}.

\subsection{X-ray model with \textsc{RefleX}}\label{sect:xraymodel}

To build a physically motivated and self-consistent X-ray model, we used \textsc{RefleX} \citep{2017A&A...607A..31P}. \textsc{RefleX} is a ray-tracing code that simulates the propagation of X-ray photons (0.1\,\,keV--1\,\,MeV), considering the most common physical processes, allowing to consider different geometries of the reprocessing material. \textsc{RefleX} tracks each photon using Monte Carlo simulations, and can be used to simulate images and spectra of AGN using quasi arbitrary geometries. This makes it possible to reproduce realistic structures of the circumnuclear material in AGN, including all different components. \textsc{RefleX} records all the events experienced by each photon, with the level of detail requested by the user. 

\begin{figure} 

\centering
\includegraphics[width=0.54\textwidth]{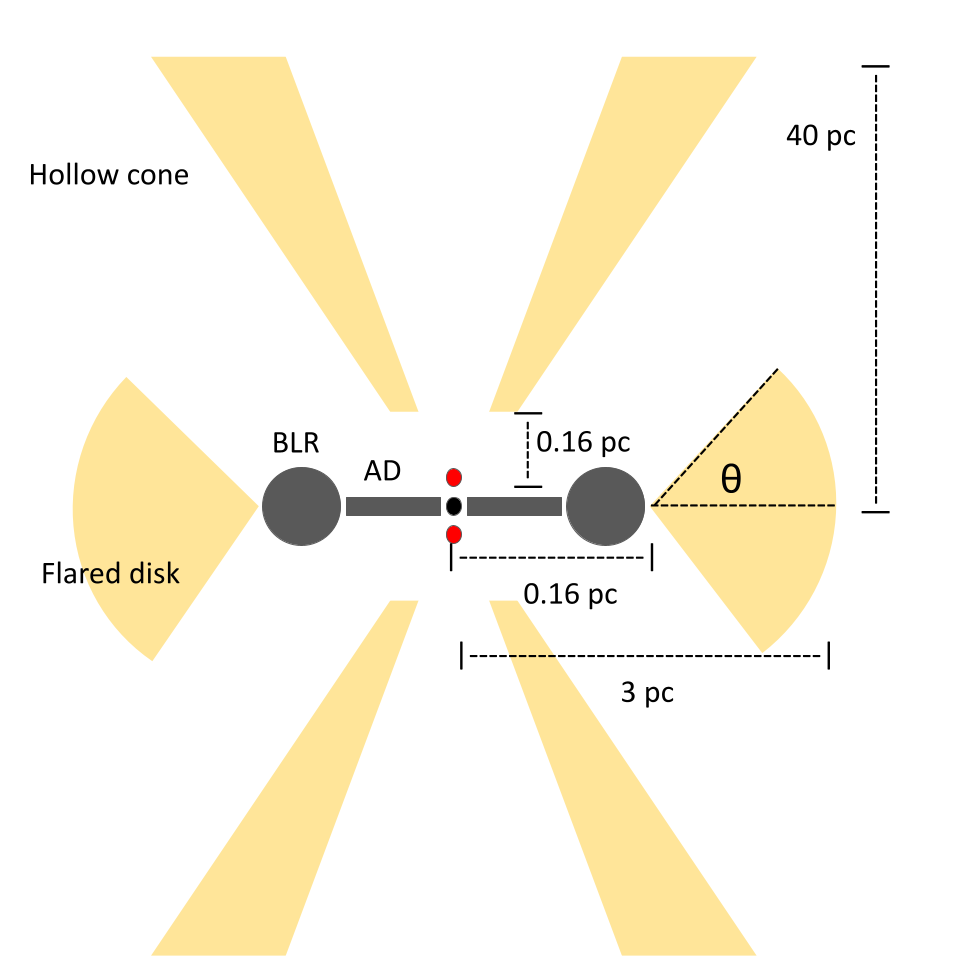}  

 \caption[]{Sketch of the X-ray model for the Circinus Galaxy. The dust-free components, i.e. the accretion disk and the BLR, are shown in grey, while the dusty components, i.e. the flared disk and the hollow cone, are shown in yellow. The red spheres represent the X-ray corona, while the central black sphere represents the SMBH, which is only put in the sketch to illustrate the center of the simulations. The size the different components are not on scale. Details on the models are reported in Table\,\ref{t:xray_model}. \label{fig:xr_model} }  
 \end{figure}

The X-ray source is simulated as a sphere with a radius of 6 gravitational radius ($ r_{\rm g} = \rm 2GM_{BH}/c^2$) and located at 10 $r_{\rm g}$ above the SMBH, following the the lamp-post geometry \citep{1996MartocchiaMatt,2004MiniuttiFabian}, and consistent with the results obtained by recent microlensing \citep[e.g.,][]{2009ApJ...693..174C} and X-ray reverberation studies \citep[e.g.,][]{2009Natur.459..540F,2013MNRAS.431.2441D} (see Figure\,\ref{fig:xr_model}). While the sources used in the previous studies are Type 1, they have an Eddington ratio similar to that of Circinus, which could imply that they share similar properties of the X-ray corona. The radius and location of the corona are equivalent to $10^{-8}\rm \: pc$ and $1.67\times 10^{-8}\rm \: pc$, respectively, considering that the mass of the black hole in Circinus is $M_{\rm BH} = 10^{6.23}\:M_{\odot}$ \citep{2017ApJ...850...74K}. The mass was estimated using the relation between $M_{\rm BH}$ and the stellar velocity dispersion from \citet{2013Kormendy}, and it is consistent with the mass inferred by \citet{2003Greenhill} from the $\rm H_2O$ maser emission.

To build the X-ray model, we include all the different dusty components of the MIR model. We make the assumption that the gas is co-spatial with the dust, and we do not consider gas depletion. X-ray photons, however, can also be reprocessed by dust-free gas, hence we included in our model the accretion disk (AD) and the BLR. A current limitation of \textsc{RefleX} is that does not include Doppler broadening due to Keplerian motion, therefore emission lines are not broadened. This feature will be implemented in a future release of the code. For the accretion disk, we adopted a flat disk. We used typical values for the properties of the accretion disk, considering recent literature. The inner radius was set to $ r_{\rm int}=\rm 6\:r_g$ and the outer radius to $r_{\rm ext}=\rm 400\:r_g$  (e.g., \citealp{2011MNRAS.414.1269W}), equivalent to $r_{\rm int}= 10^{-8}\rm\:pc$ and $ r_{\rm ext}=6.7\times 10^{-7}\rm\:pc$, respectively. The density of the disk was set to $ n_{\rm H}= \rm 10^{15}\: cm^{-3}$ (e.g., \citealp{2010ApJ...718..695G}). We considered that Hydrogen and Helium are fully ionized in the accretion disk.

To simulate the BLR we adopt a toroidal structure. Recent imaging studies carried out with Gravity have shown that the BLR could be modelled with a thick disk (e.g., \citealp{Gravity-Collaboration:2018ci,Gravity-Collaboration:2020dh,Gravity-Collaboration:2021pt}). Consistent results were obtained by dynamical modelling of the BLR using reverberation mapping data (e.g., \citealp{Pancoast:2014uv}), and by the recent micro-lensing study of \citet{Hutsemekers:2021do}. We tested whether having a thick disk or a toroidal BLR in our model would affect our results and, assuming the same covering factor and density, we found only a marginal difference between the two above $\sim 20$\,keV. We compute the inner radius of the BLR, $R_{\rm BLR}$, with the expression derived by \citet{2005ApJ...629...61K} from reverberation mapping

\begin{equation}\label{eq:r_blr}
 \frac{R_{\rm BLR}}{\rm  10\:lt\:days}=0.86^{+0.18}_{-0.15}\times \left( \frac{L_{\rm 2-10}}{\rm 10^{43}\:erg\:s^{-1}} \right)^{0.544\pm 0.091}
\end{equation} 

\noindent where $L_{\rm 2-10}$ is the X-ray luminosity in the 2-10 keV range. We use $L_{\rm 2-10}=3\times 10^{42} \rm \: erg \: s^{-1}$, i.e. the value obtained by by \citet{2014ApJ7...91...81A} from their X-ray spectral fitting. There is not a clear consensus about the covering factor of the BLR ($CF_{\rm BLR}$) in the literature. \citet{2007AJ....134.1061D} studied the far UV spectrum of 72 Seyfert galaxies and found an average covering factor of $CF_{\rm BLR} \sim 0.4$. \citet{2018MNRAS.481..533L} studied the well-known source NGC 5548 and using reverberation mapping concluded that $ CF_{\rm BLR}=0.3-0.6$. \cite{Gravity-Collaboration:2018ci} used direct imaging to find that for the quasar 3C 273 $CF_{\rm BLR}=0.7$. Based on those previous studies, we set $CF_{\rm BLR}=0.4$. The volumetric density of the gas used for the BLR is $n_{\rm H}=\rm 10^9\: cm^{-3}$, a value routinely adopted in the literature \citep[e.g.][]{1992ApJ...387...95F, 2012MNRAS.426.3086G}. We consider that the BLR is cold (i.e., all the atoms are neutral) and that the fraction of Hydrogen in molecular form is zero, following previous literature based on X-ray irradiated molecular gas \citep[e.g.,][]{1996Maloney,2019Kawamuro}.

For the flared disk, we adopt the results of S19, therefore we consider that the flared disk originates at the sublimation radius and set the outer radius to the value obtained by S19. 
To compute the equatorial column density of the cone and disk based on the optical depth of the MIR model, we use the relation reported by \citet{2014A&A...567A.142R} following \cite{2001A&A...365...28M,2001A&A...365...37M}:

\begin{equation} \label{eq:nh}
    N_{\rm H}= 1.086\: \times \tau_{\rm V} \:\times 1.1 \times \rm  10^{22} \:cm^{-2} 
\end{equation}

\noindent where $\rm \tau_{V}$ is the dust optical depth. Assuming this relation, the values of the gas column density based on the model of S17 are $ N_{\rm H,d,S17}= \rm 5.61\times10^{23}\: \rm cm^{-2} $ and $ N_{\rm H,c,S17}= \rm 1.19\times10^{22}\: cm^{-2}$, for the disc and cone, respectively, and based on the model of S19 are $ N_{\rm H,d,S19}= \rm 1.68\times10^{24}\:cm^{-2}$ and $ N_{\rm H,c,S19}= \rm 6.39\times10^{23}\:cm^{-2}$. 
The equatorial column density and the covering factor of the flared disk are free parameters in our model, having values between $ 10^{24} \:{\rm cm^{-2} }\leq  N_{\rm H,d} \leq 10^{25.5} \:{\rm cm^{-2} }$ and $ 0.1 \leq CF \leq 0.7$ respectively, consistent with S19. The choice of these ranges is also supported by previous works, which concluded that the X-ray reprocessor of Circinus is CT and has a significant covering factor \citep[e.g.,][]{2014ApJ7...91...81A}. The flared disk shape considered by S17 and S19 is not incorporated yet in \textsc{RefleX}, hence, we simulate the flared disc using a superposition of annuli that follow a radial step function that approximates the flared disk.

For the polar component we tested both the (hollow) hyperboloid and cone geometry. \textsc{RefleX} does not allow yet a hyperboloid geometry, hence we simulated it with a succession of truncated cones. We did not find any clear X-ray spectral difference between the two, therefore we adopt the cone geometry, which was computationally less expensive. The parameters of the cone are set to the values reported by S17, while the radial column density of the cone is a free parameter, allowed to vary between $ 10^{22}{\rm\,cm^{-2}}\leq N_{\rm H,c} \leq 10^{24}\rm\,cm^{-2}$, consistent with the value spredicte MIR model. While the final model of S19 has a clumpy hyperboloid shell, for simplicity and numerical limitations, here we considered only homogeneous model. We simulate the flared disk and the polar component as cold reprocessors, with a fraction of Hydrogen in molecular form set to 0.4, following previous literature \citep[e.g.,][]{2009ApJ...702...63W}.

All the components and sizes of the model are summarized in Table\,\ref{t:xray_model}, while a schematic representation of the model is illustrated in Figure\,\ref{fig:xr_model}. For each simulation, we generate $5\times 10^{8}$ photons with energies in the 0.3--70\,keV range, with a spectral binning of 10\,eV between 0.3 and 10\,keV, and 450\,eV between 10 and 70\,keV.
The photons follow a powerlaw with a cut-off energy of 200 keV \citep{Ricci:2018mp} and a photon index ($\Gamma$), which is free to vary between 1.6 and 2.4, consistent with most nearby AGN \citep[e.g.][]{Nandra:1994gu,2017ApJS..233...17R}. We include the redshift of Circinus $z=0.001448$ \citep{2004AJ....128...16K} and we consider the abundance of \citet{2003ApJ...591.1220L}. We collected all the photons between $80^{\circ}$ and $90^{\circ}$, consistent with the idea that Circinus is observed edge on \citep[e.g.][]{2003Greenhill,2014ApJ7...91...81A}. We only collect the photons in the +$z$ direction, since \textsc{RefleX} currently only allows users to incorporate one emitting source.

We generate the model by performing a large number of simulations on a grid of different parameters. The four variable parameters in our simulations are $\Gamma$, the covering factor of the flared disk ($CF$), $ N_{\rm H,c}$ and $ N_{\rm H,d}$, with a binning of 0.2, 0.1, $10^{0.2}\:\rm cm^{-2}$ and $10^{0.2}\:\rm cm^{-2}$, respectively. In total we simulate 3565 spectra, which are then used to construct the spectral model. We create separate spectra for the continuum, the scattered emission and the fluorescent emission line, and we refer to them as \textsc{model\_c} + \textsc{model\_s} +\textsc{model\_l}, respectively. We stress that the number of parameters in our model is similar to that of X-ray spectral models commonly used to fit the X-ray spectra of heavily obscured AGN.

\begin{figure} 

\hspace{-1cm}
\includegraphics[width=0.6\textwidth]{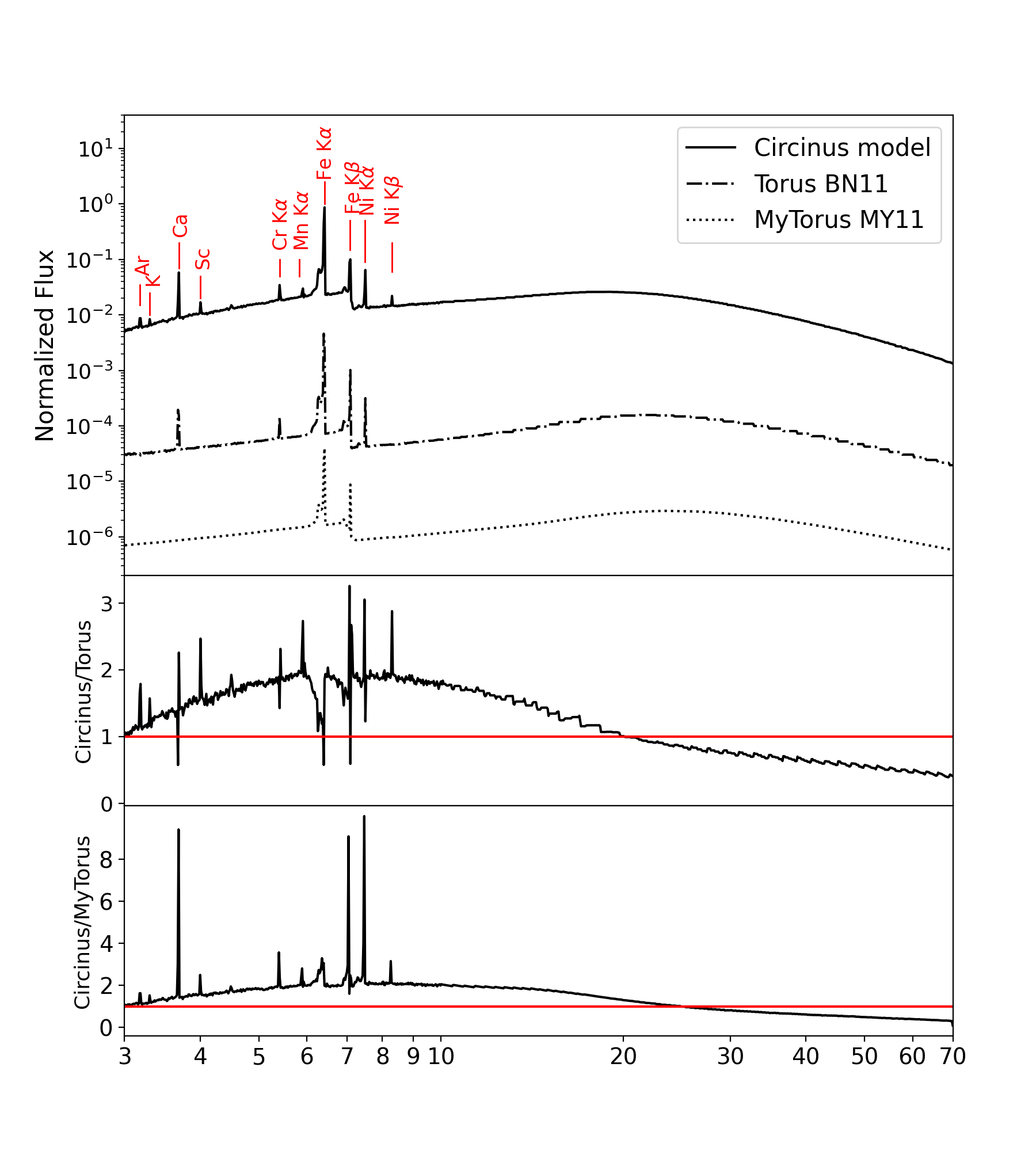}  

 \caption[]{Comparison of our model (continuous line) with the \textsc{Torus} (\citealp[][]{2011MNRAS.413.1206B},  dashdoted line) and \textsc{MYTorus} (\citealp[][]{2009MNRAS.397.1549M} dotted line) models. The fluorescent emission lines reproduced by our model are highlighted in red. The middle and bottom panels show the ratio between our model and \textsc{Torus} and \textsc{MYTorus}, respectively. The models are normalized by the 3-70 keV flux. \label{fig:model_com}}  
 \end{figure}

Our model considers several components that are typically not included in the AGN spectral models that are applied to heavily obscured AGN, such as the BLR, the polar cone, and the accretion disk. In order to show the novel features of our model, we compare it with two torus models. Figure\,\ref{fig:model_com} shows our best-fit model, together with the \textsc{Torus} \citep[][]{2011MNRAS.413.1206B} and \textsc{MYTorus} \citep[][]{2009MNRAS.397.1549M} models, obtained assuming the same properties of the torus. The figure also shows the ratio between the models, obtained after normalizing the flux. Our model produces more emission lines because it considers more fluorescent processes, which is crucial to self-consistently reproduce the emission lines that are commonly observed in X-ray spectra of AGN. For example, Mn K$\alpha$ at 5.88\,keV was identified as Cr XXIV in \citet{2014ApJ7...91...81A} using \textsc{MYTorus}, while it is naturally produced by our model. The accretion disk and the BLR can strongly affect the reprocessed continuum emission and the Fe K$\alpha$ complex, while the polar component is expected to contribute significantly to the AGN emission at $E\lesssim 5$\,keV \citet{McKaig:2021vw}, enhancing the flux of some of the emission lines (e.g., \citealp{2019MNRAS.490.4344L,McKaig:2021vw}). 

\begin{figure} 

\hspace{-1cm}
\includegraphics[width=0.6\textwidth]{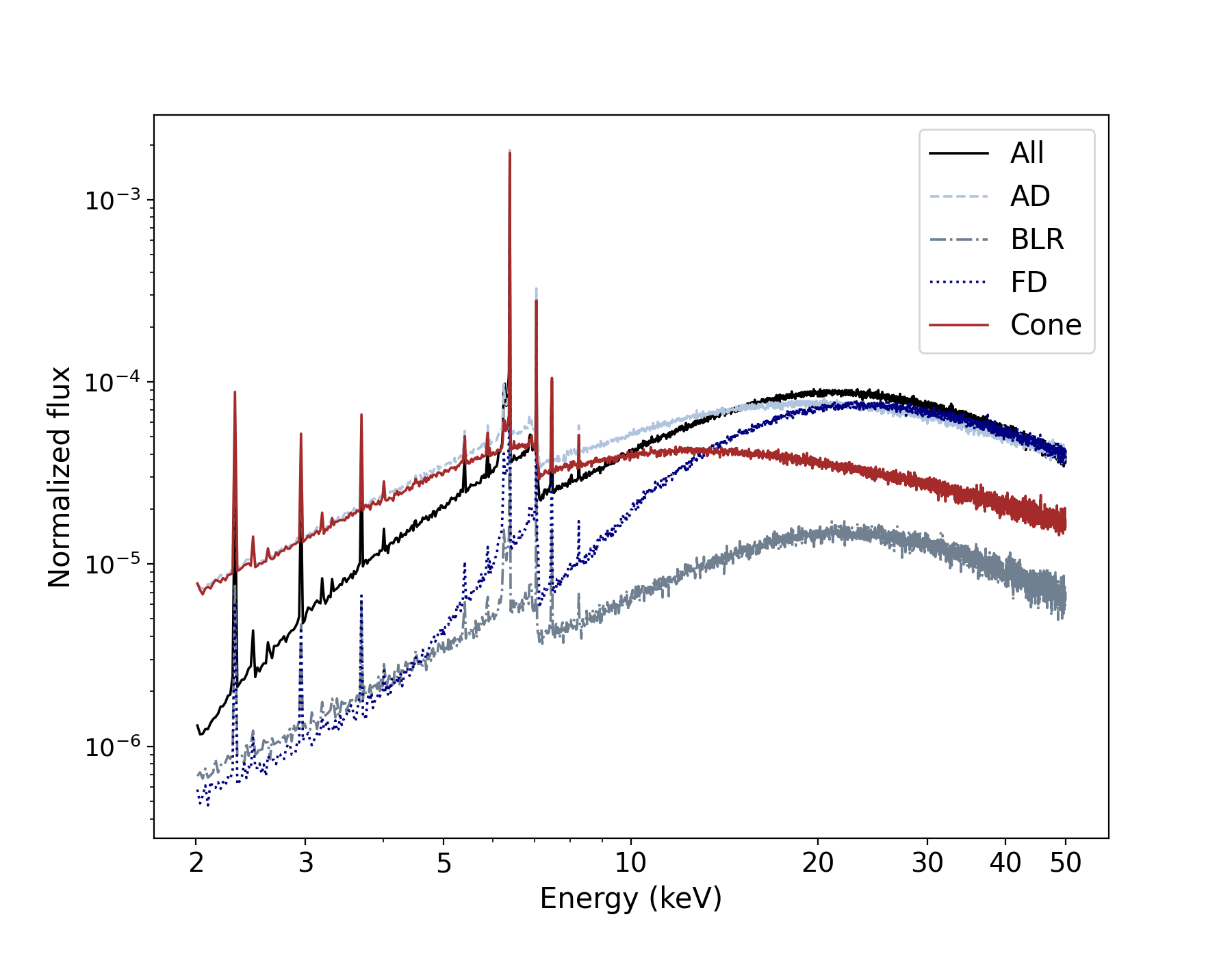}  

 \caption[]{Reprocessed radiation obtained by considering the different components of our model. The light blue, grey, blue, and red curves show the radiation from the accretion disk (AD), the broad line region (BLR), the flared disk (FD) and the shell cone, respectively. The black line shows the reprocessed radiation considering all four components, including photons that interact with more than one component. The simulations were ran using the best-fitting parameters of our model reported in Section\,\ref{sec:bb_fit} \label{fig:spec_components}.}  
 \end{figure}

To highlight the contribution of each component of the model to the final spectrum, we ran simulations of each reprocessor using the best-fitting parameters found in Section\,\ref{sec:bb_fit}. Figure\,\ref{fig:spec_components} shows the reprocessed radiation of each component of our model together with the final spectrum (i.e., including all the components in the simulation). It can be seen that the cone contributes significantly to the emission lines and to the emission at $<10$\,keV, while the flared disc and the accretion disc dominate the reprocessed radiation at high energies. The BLR contributes to the reprocessed radiation at all energies, but it never dominates the overall emission. The contribution of the accretion disc is very high due to its proximity to the X-ray corona.

\begin{table}
\centering
 \begin{tabular}{l c } 
\multicolumn{2}{c}{\bf \Large X-ray model} \\
 \noalign{\medskip}
 \hline
 \hline
 \noalign{\medskip}

\multicolumn{2}{c}{\bf \large X-ray Corona} \\
 \hline
 \noalign{\smallskip}
Radius & $\rm 10^{-8}\:pc$\\
 \noalign{\smallskip}
Height & $\rm 1.67\times 10^{-8}\:pc$\\
 \noalign{\smallskip}
Photon index ($\Gamma^*$) & [1.5; 2.4] \\
 \noalign{\smallskip}
High-energy cut-off ($E_{\rm c}$) & $200\:\rm keV$ \\
\noalign{\medskip}
 \noalign{\smallskip}

\multicolumn{2}{c}{\bf \large Accretion Disk} \\
 \hline
 \noalign{\smallskip}
Inner radius   & $\rm 10^{-8}\:pc$ \\
\noalign{\smallskip}
Outer radius   &  $\rm 6.7\times 10^{-7}\:pc$\\
\noalign{\smallskip}
Volumetric density   & $10^{15}\:\rm cm^{-3}$\\
\noalign{\medskip}
 \noalign{\smallskip}

\noalign{\medskip}
 \noalign{\smallskip}
\multicolumn{2}{c}{\bf \large Broad Line Region} \\
 \hline
 \noalign{\smallskip}
Inner radius ($R_{\rm BLR}$)  & 0.0038 pc\\
\noalign{\smallskip}
Covering factor ($CF_{\rm BLR}$) & 0.4 \\
\noalign{\smallskip}
Volumetric density   & $10^9\:\rm cm^{-3}$\\
\noalign{\medskip}
 \noalign{\smallskip}

\multicolumn{2}{c}{\bf \large Flared disc} \\
 \hline
 \noalign{\smallskip}
Covering factor ($ CF^*)$   & [$0.1;0.7$] \\
\noalign{\smallskip}
Inner radius    & 0.16 pc \\
\noalign{\smallskip}
Outer radius    & 3 pc \\
\noalign{\smallskip}
Inclination angle    & $80^{\circ}$ \\
\noalign{\smallskip}
$ N_{\rm H,d}^*$   & [$10^{24};10^{25.5}$] $\rm cm^{-2}$ \\
\noalign{\medskip}
 \noalign{\smallskip}

\multicolumn{2}{c}{\bf \large Shell cone} \\
 \hline
 \noalign{\smallskip}
Half opening angle    & $40^{\circ}$ \\
\noalign{\smallskip}
Angular full width    & $10^{\circ}$ \\
\noalign{\smallskip}
Inner height   & 0.16 pc \\
\noalign{\smallskip}
Radial extension   & 40 pc \\
\noalign{\smallskip}
$  N_{\rm H,c}^*$   & [$10^{22};10^{24}$] $\rm cm^{-2}$  \\
\noalign{\medskip}
 \noalign{\smallskip}

  \hline
 \noalign{\smallskip}
\end{tabular}
 \caption{Parameters of the X-ray model, including all the components: the X-ray corona, the Accretion Disk, the Broad Line Region, the flared disk and the shell cone. $ N_{\rm H,d}$ and $ N_{\rm H,c}$ are the equatorial column density of the flared disk and the radial column density of the shell cone, respectively. These are the maximum column densities of the polar and equatorial components. The free parameters of the model are identified with $^*$. Further details can be found in Sect.\,\ref{sect:xraymodel}. }
 \label{t:xray_model}
\end{table}

\section{X-RAY OBSERVATIONs AND DATA REDUCTION} \label{sec:xray} \label{sect:data_reduction}

We apply our model to the X-ray observations of Circinus carried out by \textit{NuSTAR} (for a total exposure of  53.4\,ks) and \textit{Chandra} (671.4\,ks). The data analysis of the \textit{NuSTAR} observations is already reported in \citet{2014ApJ7...91...81A}. For further details about the observations, data reduction and the spectral extraction, we refer the readers to Section\, 2 of \citet{2014ApJ7...91...81A}. The \textit{Chandra} and \textit{NuSTAR} observations analyzed in this paper are listed in Table \,\ref{t:data}.

\subsection{\textit{NuSTAR}} \label{sub:nustar}

We use here the \textit{NuSTAR} \citep{2013ApJ...770..103H} observations of Circinus performed in 2013-01-25. We consider 100\farc{} radius aperture \textit{NuSTAR} spectra for modules A and B. The spectra are contaminated by 21 point sources, where the brightest are two point sources detected in the \textit{XMM-Newton} observations: the X-ray binary (XRB) CGX1 and the supernova remnant CGX2 (see Figure\,1 of \citealp{2014ApJ7...91...81A}). Both sources were originally reported by \citet{2001AJ....122..182B}.

\subsection{\textit{Chandra}}

Circinus was observed several times by \textit{Chandra} \citep{2000SPIE.4012....2W} ACIS and HETG (see Table\,\ref{t:data} for details of each observations). The ACIS-I and ACIS-S CCD-resolution spectra are discarded because they are highly affected by pile-up \citep{2014ApJ7...91...81A}, and we only consider the HETG data for this work, using the spectra from the High Energy Grating (HEG, 0.8--10.0 keV). \citet{2014ApJ7...91...81A} did not find any variablity in the different observations, therefore we combine all the spectra.

We reduce the observations using the standard \texttt{chandra\_repro} script, within the CIAO software (v 4.11), and recent calibration files (CALDB v 4.8.3). We used the \texttt{tg\_create\_mask} and \texttt{tg\_resolve\_events} tasks to identify the HEG arms and to resolve the different spectral orders. Then, we used \texttt{tgextract} to extract aperture-corrected spectra. We extract the 1st-, 2nd-, and 3rd-order spectra using a range of extraction boxes in order to assess whether any portion of the nuclear emission was resolved. Among these, we use the 3-pixel spectra to model the unresolved AGN emission, as this aperture is well-matched to the nominal spatial resolution of Chandra (80$\%$ encircled energy radius is 0.69\farc{}). While 2nd- and 3rd-orders have smaller effective areas, they offer higher spectral resolution compared to the 1st order, which can help to determine whether the emission lines are resolved or not. The larger 12-pixel aperture spectra are used to perform a broadband fit, as it captures the total nuclear emission with smaller uncertainties due to minimal aperture corrections.

The spectra are not grouped. The majority of the channels have $\sim 1$ count, and the counts per bin reach a maximum of $\sim 40$ around the Fe K$\alpha$ line. We did not rebin the spectrum in order to avoid losing any spectral information. Hence, the fits are carried out using Cash statistics \citep{1979Cash}.

To model the spectra of Circinus it is necessary to model all the individual point sources included in the \textit{NuSTAR} aperture. We use the contamination spectra extracted and modelled by \citet{2014ApJ7...91...81A}. They used the ACIS-S CCD spectra (OBSIDs 12823 and 12824) and extracted 4\farc{} aperture spectra for each contaminating source.   

\begin{table}
\centering
 \begin{tabular}{l c c c } 
 \hline
 \hline
 \bf Observatory &\bf Date&\bf Exposure [ks]&\bf OBSID \\
 \hline
 \noalign{\smallskip}

 \textit{Chandra} HETG & 2000-06-15 & 68.2 & 374+62877\\
 \noalign{\smallskip}
 \textit{Chandra} HETG &2004-06-02 &55 &4770 \\
 \noalign{\smallskip}
 \textit{Chandra} HETG & 2004-11-28&59.5 & 4771 \\
  \noalign{\smallskip}
 \textit{Chandra} HETG &2008-12-08 &19.7 & 10226\\
  \noalign{\smallskip}
 \textit{Chandra} HETG &2008-12-15 &103.2 & 10223\\
  \noalign{\smallskip}
 \textit{Chandra} HETG &2008-12-18 & 20.6&10832 \\
  \noalign{\smallskip}
 \textit{Chandra} HETG &2008-12-22 & 29.0& 10833\\
  \noalign{\smallskip}
 \textit{Chandra} HETG & 2008-12-23& 77.2& 10224\\
 \noalign{\smallskip} 
 \textit{Chandra} HETG & 2008-12-24&27.8 &10844 \\
  \noalign{\smallskip} 
 \textit{Chandra} HETG &2008-12-26 & 68.0& 10225\\
  \noalign{\smallskip}
 \textit{Chandra} HETG & 2008-12-27&37.4 &10842 \\
  \noalign{\smallskip}
 \textit{Chandra} HETG &2008-12-29 & 57.3& 10843\\
  \noalign{\smallskip}
 \textit{Chandra} HETG &2009-03-04 & 16.5& 10872\\
  \noalign{\smallskip}
 \textit{Chandra} HETG & 2009-03-03& 13.9& 10850\\
   \noalign{\smallskip}
 \textit{Chandra} ACIS-S & 2010-12-17& 152.4& 12823\\
  \noalign{\smallskip}
\textit{Chandra} ACIS-S & 2010-12-24& 38.9& 12824\\
\noalign{\smallskip}
\textit{NuSTAR} & 2013-01-25 & 53 &60002039002 \\ 
 \noalign{\smallskip}
\textit{NuSTAR} & 2013-02-02 & 18 &30002038002 \\
 \noalign{\smallskip}
\textit{NuSTAR} & 2013-02-03 & 53 &30002038004 \\
 \noalign{\smallskip}

\textit{NuSTAR} & 2013-02-05 & 36 & 30002038006 \\

\hline
 \end{tabular}
 \caption{Summary of the observations used in our work.} \label{t:data}
\end{table}

\section{X-ray spectral fitting}\label{sec:xrayspectralfitting}

We fit the X-ray data using \textsc{xspec} V12.10.1f, \citep{1996ASPC..101...17A}. The \textit{NuSTAR} spectra are modeled considering both the emission from the AGN and the contaminating extra-nuclear point sources and off-nuclear reflection. To fit the contamination, we use the model implemented by \citet{2014ApJ7...91...81A}. For the AGN we used our specific X-ray model built for Circinus, in which the different components, i.e. transmitted, scattered and the emission lines, are distinct table models (see Sect.\ref{sect:xraymodel} for details). We tie the normalization of the transmitted and scattered component, while leaving free to vary the normalization of the emission lines component. We also include several Gaussian lines reported by \citet{2001ApJ...546L..13S} to model the colder, photo-ionized medium, associated with radiative recombination continuum and line emission (hereafter RRC and RL, respectively, following \citealp[]{2015Bauer}) which is not included in our \textsc{RefleX} model. The equivalent width and the normalization of the lines are left free to vary. Table\,\ref{t:lines} of the Appendix shows the best-fit parameters of the lines.

\subsection{Modelling the nuclear spectrum with \textit{Chandra}} \label{sec:chandra_fit}

\begin{figure*} 

\centering
\includegraphics[trim=10 0 210 20,clip,width=0.85\textwidth]{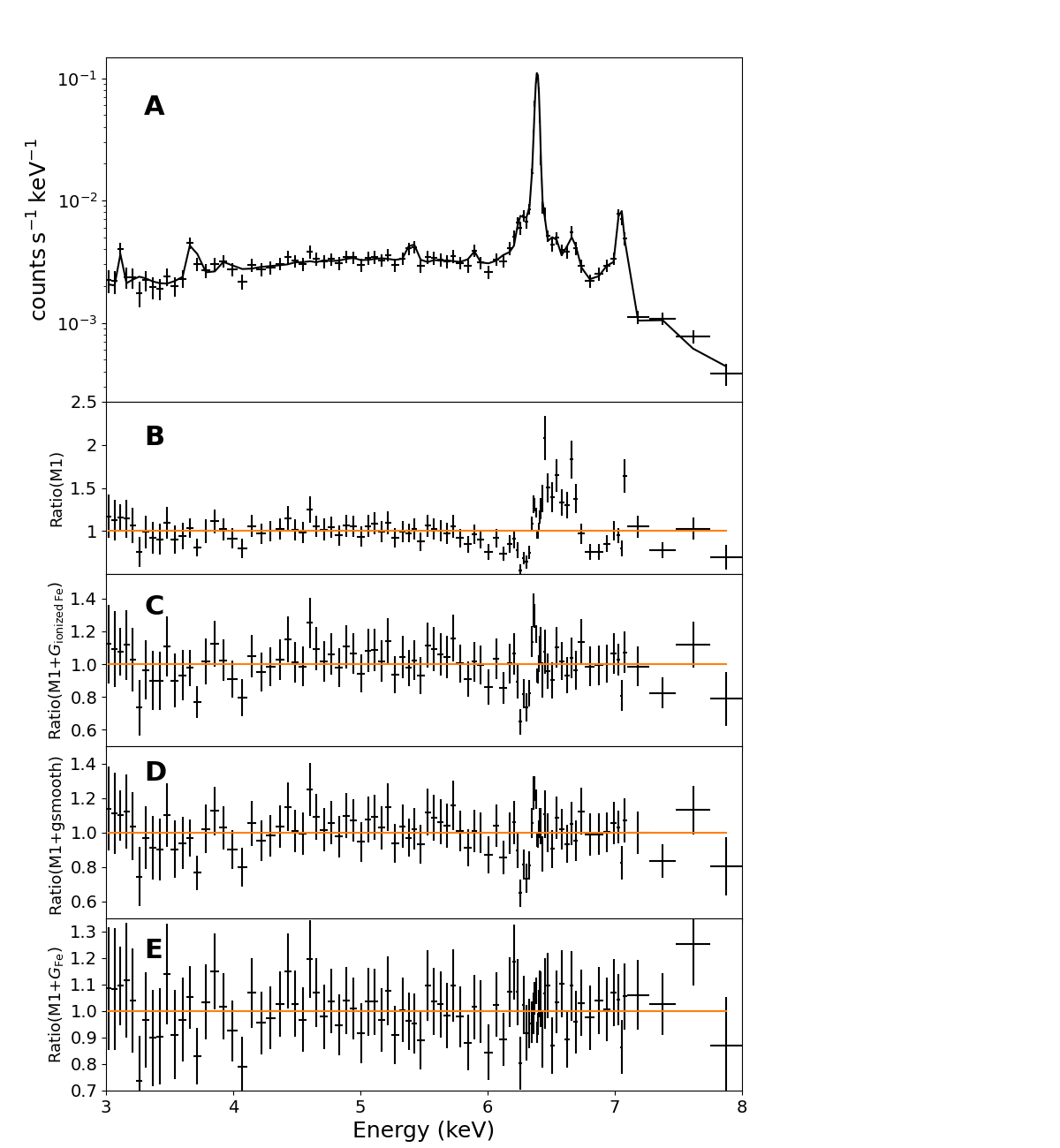}  

 \caption[]{Best-fitting models for the central 3\farc{} \textit{Chandra}/HEG spectrum. Panel A shows the observed spectra together with our best-fitting model \textsc{M1} + \textsc{G}$_{\rm Fe}$, which consider our \textsc{RefleX} model plus Gaussian emission lines for Fe K$\alpha$, Fe K$\beta$, and ionized Fe. Panel B shows the ratio between the data and the model \textsc{M1}, which is our \textsc{RefleX} model. Panel C shows the ratio between the data and the model \textsc{M1}+\textsc{G}$_{\rm ionized Fe}$, which also includes Gaussian emission lines to reproduce the ionized iron lines. Panel D represents the ratio between the data and the model \textsc{M1}+gsmooth, which is our \textsc{RefleX} model but considering that the BLR emission is affected by Doppler broadening, and also Gaussian emission lines to reproduce the ionized iron lines. Lastly, panel E shows the ratio of the data with the model \textsc{M1} + \textsc{G}$_{\rm Fe}$. The Gaussian emission lines added to the fits are reported in Table\,\ref{t:FeK_complex}. For more details of the models, see Section\,\ref{sec:chandra_fit}.\label{fig:heg_fit} }  
 \end{figure*}

We start by fitting the 3--8 keV nuclear spectrum using the high-resolution data from \textit{Chandra}/HEG. The spectrum (see panel A of Figure\,\ref{fig:heg_fit}) shows a large number of narrow emission lines at low energies, and strong features at 6.4, 7.05 and 7.4 keV, corresponding to neutral Fe K$\alpha$, Fe K$\beta$ and Ni K$\alpha$, respectively. As discussed by \citet{2001ApJ...546L..13S}, the emission lines in Circinus come from two different regions: nuclear neutral material (i.e., central few parsecs) and ionized material located farther away from the SMBH (RRC and RL), likely related to the narrow line region \citep{2007MNRAS.374.1290G}. An important feature in the spectrum is the Fe K$\alpha$ Compton shoulder (CS). The CS is created when Fe K$\alpha$ photons are Compton-down scattered, which produces a low-energy tail associated with the line \citep[e.g.,][]{1991MNRAS.249..352G,1991A&A...247...25M}. The observation of the CS indicates that the reflecting material has a very high column density and/or a high metal abundance (e.g., \citealp{2016MNRAS.462.2366O}).

\begin{table*}
\centering

 \begin{tabular}{lcccc} 
 \hline
 \hline
\noalign{\smallskip}
(1) &  (2) & (3) & (4) & (5) \\
\noalign{\smallskip}
\bf Line &\bf  Energy &{\bf Width} & {\bf $v_{\rm FWHM}$}&{\bf Normalization} \\
\noalign{\smallskip}
 &  [keV] &[eV] & $(\rm km\: s^{-1})$ & $(\rm 10^{-4} photons\: cm^{-2}\: s^{-1})$\\
\noalign{\smallskip}
 \hline 
\noalign{\smallskip}
 Fe K$\alpha$ & $6.396\pm 0.001$ &$13\pm 1.4$& $1346\pm 155$ & $1.38 \pm 0.1 $\\
 Fe K$\alpha$ IV-V &$6.456\pm 0.006$ & $<5 $ & <547 & $0.083 \pm 0.03$  \\
 Fe K$\alpha$ VI-XII &$6.53\pm 0.012$ & $30_{-5.2}^{+ nc} $ & $3242^{+ nc}_{-562}$ & $0.14 \pm 0.031$\\
 Fe K$\alpha$ XXV & $6.671\pm 0.011$ & $36^{+ nc}_{-25}$ & $3812^{+ nc}_{-2647}$ & $ 0.13\pm 0.03$\\ 
  Fe K$\alpha$ XXVI & $6.973$ & $<15$ & $<1520$ & $ 0.13\pm 0.03$\\ 
 Fe K$\beta$ & $7.05\pm 0.007$ & $13\pm 1.4$ & $1346\pm 155$ & $0.18 \pm 0.04 $ \\
 Fe K$\beta$ II & $7.09\pm 0.009$ & $<5$ & <498 & $0.09 \pm 0.03 $ \\
 
\noalign{\smallskip}
 \hline
 \end{tabular}
 \caption{ Best-fit parameters for the Gaussian emission lines found around the Fe K$\alpha$ complex. The different columns show the: (1) line (2) centroid energy of each Gaussian, (3) line widths, (4) Full Width at Half Maximum, and (5) normalization of the Gaussian lines. $nc$ implies that the parameter was not constrained.} \label{t:FeK_complex}

\end{table*}

\begin{figure}
\hspace{-0.8cm}
\includegraphics[trim=0 0 0 0, clip,width=0.55\textwidth]{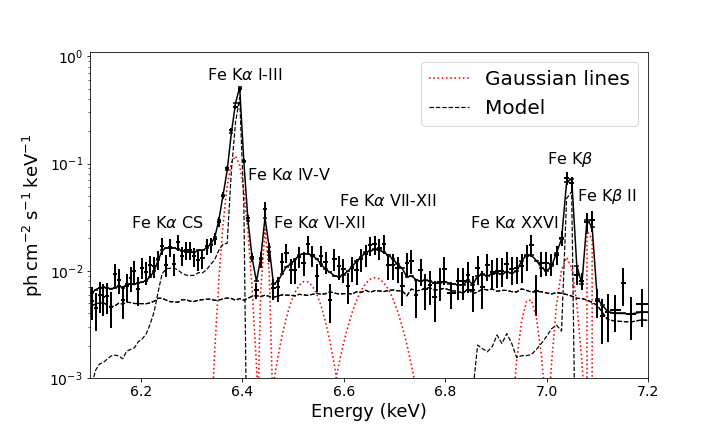}
\caption{Unfolded model in the 6.1--7.2 keV range. The black points with error bars represent the \textit{Chandra}/HEG spectrum, the black line is the total model, the black dashed lines are the scattered (\textsc{model\_s}) and the emission lines (\textsc{model\_l}) components of the model, and the red dotted lines are the additional Gaussian lines listed in Table\,\ref{t:FeK_complex}.   }\label{fig:fe_eeuf}	
\end{figure}

\begin{table*}
\centering
 \begin{tabular}{lccc} 
 \hline
 \hline
\noalign{\smallskip}

\bf Component &\bf  Parameters &{\bf M1 + \textsc{G}$_{\rm Fe}$} &{\bf M2} \\
\noalign{\smallskip}
 \hline
 \noalign{\smallskip}
\noalign{\smallskip}
 Cone model&$\Gamma$&$2.1\pm 0.11$&$1.65^{+0.11}_{-0.03}$ \\
 \noalign{\smallskip}
&$  N_{\rm H,d}$ ($\rm cm^{-2}$)&$\rm 1.62^{+0.29}_{-0.16}\times 10^{24} $&$\rm 1.01^{+0.03}_{-0.24}\times 10^{25} $  \\
 \noalign{\smallskip}
&$ N_{ \rm H,c}$ ($\rm cm^{-2}$)&$\rm 10^{+nc}_{-0.67}\times 10^{23} $&$\rm 2.18^{+0.47}_{-0.43}\times 10^{23} $ \\
 \noalign{\smallskip}
&$CF$&$0.4^{+0.03}_{-0.07}$&$0.55^{+0.01}_{-0.05}$ \\
 \noalign{\smallskip}
&$N_l$ ($\rm ph\: keV^{-2}\: s^{-1}\: cm^{-2}$) & $0.2^{+0.03}_{-0.01}$  &$0.16\pm 0.16$  \\
 \noalign{\smallskip}
 &$N_s$ ($\rm ph\: keV^{-2}\: s^{-1}\: cm^{-2}$)&$0.23 \pm 0.14 $&$0.17 \pm 0.009$   \\
 \noalign{\smallskip}
Scattered pl& $N/N_c$ &$3.1 \pm 0.5\times 10^{-3}$&$ 6 \pm 0.4\times 10^{-3}$  \\

\noalign{\smallskip}
\hline
\noalign{\smallskip}
$\rm stats/dof$ && 1104/1033 &2359.82/2397 \\
 \noalign{\smallskip}
\hline
 \noalign{\smallskip}
 \end{tabular}
 \caption{Parameters of the best-fitting models obtained for the \textit{Chandra}/HEG spectrum (\textsc{M1} + \textsc{G}$_{\rm Fe}$), and for the \textit{Chandra}/HEG and {\it NuSTAR} (\textsc{M2}; see Sections\,\ref{fig:heg_fit} and\,\ref{fig:overall_fit}) . $\Gamma$ is the photon index, $ N_{\rm H,d}$ is the equatorial column density of the flared disk,  $ N_{\rm H,c}$ is the radial column density of the cone shell, and $CF$ is the covering factor of the flared disk. $N_l$ and $N_s$ are the normalizations of the emission lines  and the scattered component of the AGN, respectively. $N/N_c$ is the ratio between normalizations of the scattered powerlaw and the continuum. The normalization of the continuum is tied to the normalization of the scattered emission ($N_c=N_s$). $nc$ means not constrained.  } \label{t:heg_fits}

\end{table*}

For the initial fit, we assume an uniform reprocessor where the obscurer and the scatterer material are the same, so the continuum (\textsc{model\_c}), Compton-scatted (\textsc{model\_s}) and emission lines (\textsc{model\_l}) components have the same $\Gamma$, $N_{\rm H,d}$ and $CF$. However, we also consider that emission from the reflected and emission lines components could originate at different scales, so we tie the normalization of the continuum $N_c$ to be the same as that of the scattered component $N_s$, while letting free the normalization of the fluorescent emission lines $N_l$ .  The model in \textsc{xspec} is:
\newline

\textsc{M1} = \textsc{phabs} $\times$ (\textsc{model\_l} + \textsc{model\_s} +\textsc{model\_c}+ \textsc{powerlaw} + 5 $\times$ \textsc{zgauss})
\newline

\noindent The \textsc{phabs} component takes into account Galactic absorption in the direction of Circinus ($ N_{\rm H}=5.3\times 10^{21}\: \rm cm^{-2} $; \citealp{2005A&A...440..775K}), the \textsc{powerlaw} reproduces the Thomson scattered emission from larger-scale material (e.g., \citealp{2007ApJ...664L..79U,2017ApJS..233...17R,2021Gupta}), and we use \textsc{zgauss} for the 5 RRC and RL Gaussian emission lines that we detect in the spectrum (see Table\,\ref{t:lines} in the Appendix). With this model, we obtain a $\rm stats/dof=2789/1859=1.5$, where $\rm stats$ corresponds to the total fit statistic (i.e., $\chi^2$ statistic from the NuSTAR and contamination spectra plus cstat statistic from the HEG spectra) and $\rm dof$ are the degrees of freedom. The ratio between the data and the model is shown in the panel B of Figure\,\ref{fig:heg_fit}. Overall, the model reproduces well the spectrum, but significant residuals are found around the Fe K$\alpha$ complex, suggesting that Circinus has emission lines associated with iron with different ionization states. 
Also, the model does not reproduce the low-energy side of the Fe K$\alpha$ line and overpredicts the CS. Therefore, we add five additional Gaussian lines to model the ionized iron lines at 6.45, 6.55, 6.67, 6.97 and 7.09 keV. These emission lines can be associated with Fe K$\alpha$ IV-V, Fe K$\alpha$ VII-XII, Fe K$\alpha$ XXV, Fe XXVI, and Fe K$\beta$ II \citep{1985Johnson,1993Kaastra}, respectively. We tie the line widths of the Fe K$\alpha$ and Fe K$\beta$ lines, as they are believed to originate from the same material, and leave free to vary the centroids, the line widths and the normalizations of the Gaussians for the rest of the emission lines. The fit parameters of the lines are listed in Table\,\ref{t:FeK_complex}. We call this new model \textsc{M1}  + \textsc{G}$_{\rm ionized \: Fe}$ and the residuals are shown in the panel C of Figure\,\ref{fig:heg_fit}. With this model, the fit substantially improves, and we obtain a $\rm stats/dof=1198/1033=1.16$. However, there are still a significant residuals around the CS. 

Previous works have found that in many AGN an important fraction of the Fe K$\alpha$ line originates in the BLR \citep[e.g.][]{2010ApJS..187..581S,2015ApJ...812..113G}, implying that the line can be Doppler broadened. We assess this possibility by applying a gaussian smoothing to the fraction of the Fe K$\alpha$ line that originates in the BLR. To achieve this, we create two new models, one considering the accretion disk, the flared disk and the hollow cone, and the other one considering just the BLR. The model in XSPEC is:  

\textsc{M1+gsmooth} = \textsc{phabs} $\times$ (\textsc{model\_l\_blrfree}+\textsc{model\_s\_blrfree}+\textsc{model\_c\_blrfree} + \textsc{gsmooth}$\times$(\textsc{blr\_l}+\textsc{blr\_s}+\textsc{blr\_c})+\textsc{powerlaw} + 11 $\times$ \textsc{zgauss})
\newline

\noindent where \textsc{model\_l\_blrfree}, \textsc{model\_s\_blrfree}, \textsc{model\_c\_blrfree}, are the emission lines, scattered component and continuum emission from the AD, the flared disk and the hollow cone, respectively. Similarly, the components \textsc{blr\_l}, \textsc{blr\_s}, \textsc{blr\_c}, are the emission lines, scattered and continuum emission of the BLR, respectively. The \textsc{gsmooth} component is broadening the BLR emission. We tie the normalization of the emission lines, scattered and continuum emission models of the BLR to the "BLR free" models. The residuals of the new model are shown in the panel D of Figure\,\ref{fig:heg_fit}. The model slightly improves, yielding to a $\rm stats/dof=1178/1033=1.14$. We find that the 15$\%$ of the Fe K$\alpha$ flux comes from the BLR, and the best fit value for the width of the Gaussian is $15_{+4.5}^{-4.8}$ eV, equivalent to $1573^{-541}_{-500} \: \rm km \: s^{-1}$, characteristic velocity of a narrow emission line. This result suggests that the Doppler broadening is not the main responsible for the residuals.

\citet{2014ApJ7...91...81A} found that there is an overall broadening of the Fe K$\alpha$ and Fe K$\beta$ lines due to Fe spatial extension, which causes contamination of the nuclear spectrum from neighboring energy bins, due to incorrect spatial vs. energy assignments along the dispersion direction. This a possible reason for the difficulty in reproducing the Fe K$\alpha$ complex of Circinus. Therefore, following the same procedure that \citet{2014ApJ7...91...81A}, we add two additional Gaussian emission lines at 6.4 and 7.05 keV. We call this new model \textsc{M1}  + \textsc{G}$_{\rm Fe}$ and the residuals are shown in the panel E of Figure\,\ref{fig:heg_fit}. Figure\,\ref{fig:fe_eeuf} illustrates the best fit model between 6.1 and 7.3 keV, with the different components of the spectrum. With this model, the fit substantially improves, and we get a $\rm stats/dof =1095/1033=1.06$, supporting the idea of a potential spatial extension of the Fe K$\alpha$.

\begin{figure}
\centering
\includegraphics[trim=0 0 0 0, clip,width=0.45\textwidth]{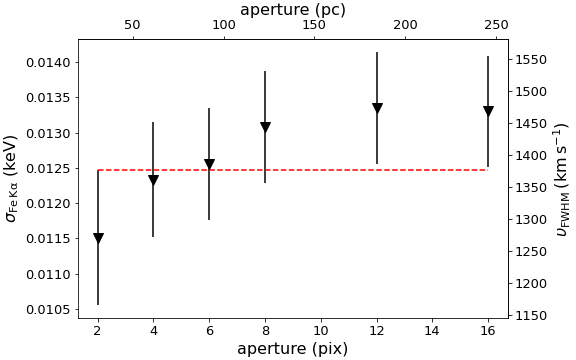}
\caption{$\rm Fe\:K\alpha$ line width ($\sigma_{\rm Fe\:K\alpha}$) against spectral aperture extraction. The lower x-axis shows the aperture extraction in pixels, the upper x-axis shows the aperture in parsec scales, the left y-axis shows the line width in energy, and the right y-axis shows the line width values converted to velocities. The red dotted line shows the upper limit value of $\sigma_{\rm Fe\:K\alpha}$ for the 2-pixel aperture spectrum. The errors are quoted at 90$\%$ confidence level. }\label{fig:width_ap}	

\end{figure}

To explore the possibility that the Fe K$\alpha$ emission is extended, we fit six 1st-order HEG spectra, with extraction apertures radius between 2 and 16 pixels. We use a simple model, consisting of a power law plus a Gaussian at 6.4 keV. The results are shown in Figure\,\ref{fig:width_ap}. We find that the line width systematically increases as the spectral aperture increases, and the line widths of the spectra with an aperture extraction higher than 12 pixels are higher than the line width for the 2-pixel aperture spectrum at the 90$\%$ confidence level. However, the line widths are consistent at a 3-$\sigma$ level. Additionally, we compare the line widths of the 1st-, 2nd-, and 3rd-order 2-pixel aperture spectra (see Figure\,\ref{fig:width_ord} in Appendix \ref{ap:emission_lines}), and find that the values are consistent, although the best fit to the 2nd-order width appears $\sim 30-40\%$ lower at >$90\%$ confidence. Taken together, this suggests that the true nuclear width may only accounts for 50--75$\%$ (see Figure\,\ref{fig:width_ap}) of the total measured in the 3 pixel aperture, implying that the Fe K$\alpha$ emission may extend up to hundreds of parsecs, although a combination of higher S/N, higher spectral resolution, and higher spatial resolution data are likely necessary to draw a firm conclusion.

We consider the last model as our best fit. Table\,\ref{t:heg_fits} shows the best fit parameters of this model. As expected, we obtained a Compton-thick column density for the flared disk of $ N_{\rm H,d}= \rm 1.62^{+0.29}_{-0.16}\times 10^{24} \: cm^{-2}$, consistent with previous estimates \citep[e.g.,][]{2014ApJ7...91...81A,2019A&A...629A..16B,2021Uematsu}, with a covering factor of $CF=0.4^{+0.03}_{-0.07}$. 

\begin{figure}
\centering
\includegraphics[trim=0 0 0 0, clip,width=0.5\textwidth]{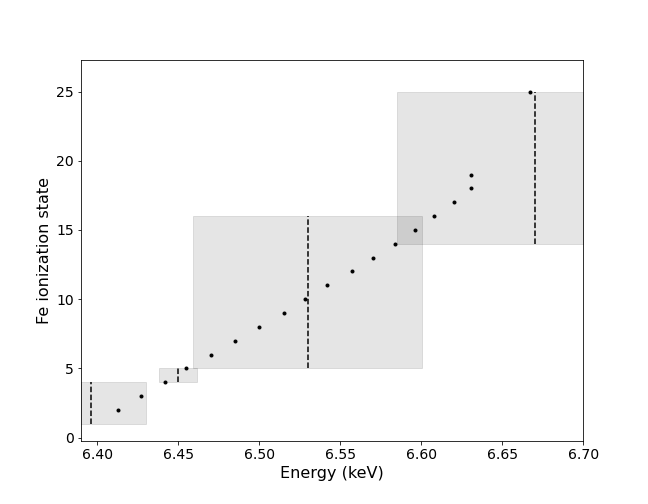}
\caption{Energy centroids of the ionized iron lines in the 6.4-6.7 keV range versus the iron ionization state (black dots). The black dashed lines represent the energy centroids of the lines detected in the Circinus spectrum, and the grey shaded areas represent the FWHM of each line.}\label{fig:ion_fe}	
\end{figure}

Table\,\ref{t:FeK_complex} lists all the ionized iron lines we detect in the Circinus spectrum. Figure\,\ref{fig:ion_fe} shows the energy centroids of all the ionized iron lines in the 6.4-6.7 keV range as a function of the iron ionization state. The black dashed vertical lines represent the energy centroids of the lines we detected in Circinus, and the shaded areas represent the FWHM of those lines. The Gaussian lines we add are covering a very broad range of ionization states in the 6.4-6.7 keV band. Nevertheless, the energy resolution of HEG does not allow to resolve each of the ionized iron lines, therefore, each of the Gaussians we add could be including emission from more than one ionized iron line. We will be able to resolve each line with the advent of {\it XRISM} \citep{2020arXiv200304962X}, a mission that will provide high energy resolution spectra.  Based on the energy centroids and line widths of the gaussian lines, we estimate the different ionized emission lines that each Gaussian component is fitting (see Table\,\ref{t:FeK_complex}). Figure\,\ref{fig:fe_eeuf} shows that the Fe XXVI line at 6.97 keV is in the same energy range as the CS of the Fe K$\beta$ line, which could lead to some degeneration between the column density of the reprocessor and the parameter of the Gaussian line. The clear excess around 6.97 keV indicates the existence of this line. Figure\,\ref{fig:fe_eeuf} teaches us the complexity of the Fe K$\alpha$ complex in Circinus and the importance of taking into account the different components of the spectra.

\subsection{Broadband spectrum} \label{sec:bb_fit}

\begin{figure} 
\centering
 \includegraphics[trim=0 160 0 40, clip,width=0.48\textwidth]{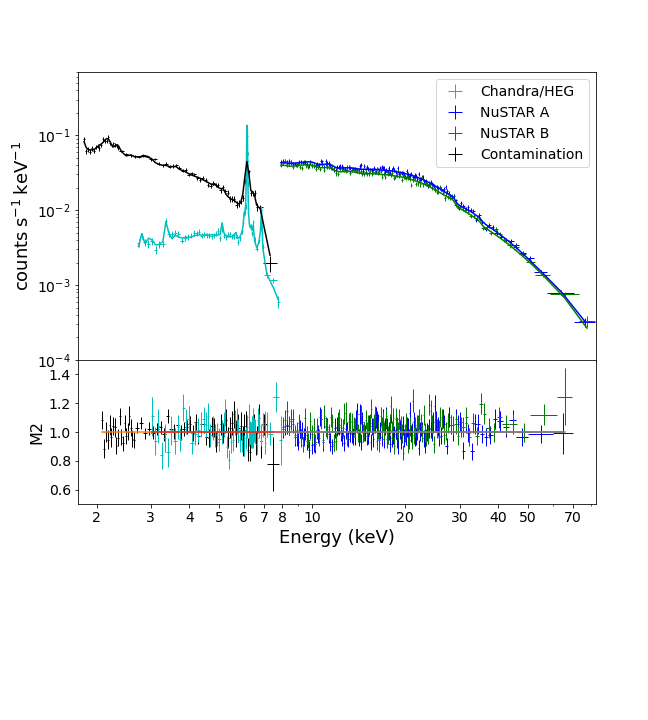} 
  \includegraphics[width=0.5\textwidth]{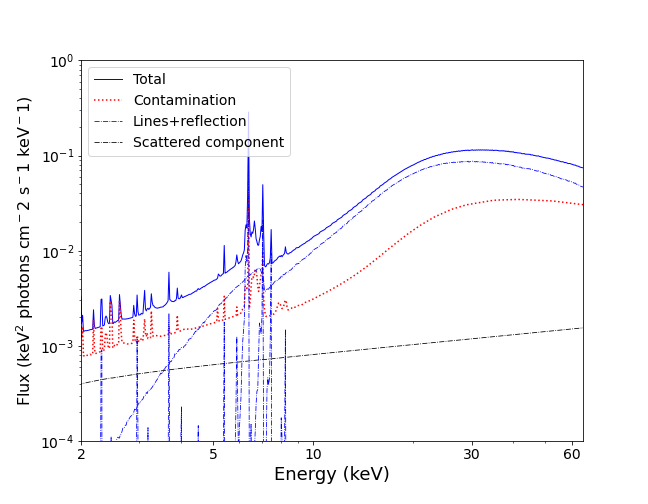} 

 \caption[]{\textit{Top figure}: \textit{NuSTAR} FPMA and FPMB spectra (blue and green, respectively), the \textit{Chandra}/HEG spectrum (cyan) and the contamination spectrum (black), altogether with the best fitting model. The bottom panel shows the ratio between the data and \textsc{model\_2}. \textit{Bottom figure}: best-fitting model for the broadband spectrum. The blue dash-dotted lines represent the emission lines and Compton scattering component of the \textsc{RefleX} model, the red dash-dotted line the scatter component at larger scales, and the red dotted line represent the contaminating sources. \label{fig:overall_fit} }  
 \end{figure}

We then extend our model to fit the spectra of both {\it Chandra} and \textit{NuSTAR} in the 8--70 keV band. We do not consider the 3--8 keV \textit{NuSTAR} spectrum as that band is affected by CGX1, which is strongly variable \citep[][]{2001AJ....122..182B}, making difficult to perform a good joint fit between the contamination and the \textit{NuSTAR} spectra. Hence, we prefer to model the low energy spectrum of Circinus using the clean nuclear data. As discussed in Section\,\ref{sub:nustar}, the \textit{NuSTAR} spectra are contaminated by 21 point sources; we included these contaminating sources using the same modelling adopted by \citeauthor{2014ApJ7...91...81A} (\citeyear{2014ApJ7...91...81A}; for more details see sections\,4.2 and\,4.4 of their paper). The top panel of Figure\,\ref{fig:overall_fit} shows the contamination spectrum (black points), {\it NuSTAR} FPMA and FPMB spectra (blue and green points, respectively) and the \textit{Chandra}/HEG spectrum (cyan points). We tested the same model as for the nuclear component,  \textsc{M1} + \textsc{G}$_{\rm Fe}$ (see Section\,\ref{sec:chandra_fit}), but now including the contaminating sources. The \textsc{contamination} model consists of a Compton-scattered continuum component to account for the off-nuclear reflection modeled by \textsc{mytorus} \citep{2012Yaqoob}, an absorbed powerlaw component with $\Gamma=1.8$ to model the point-sources emission, one powerlaw to model the scattered emission, and a thermal plasma component to reproduce the thermal emission of the diffuse region \citep{2014ApJ7...91...81A}. \newline
The model we use is: 
\newline

\textsc{M2} = \textsc{constant} $\times$ \textsc{phabs} $\times$ (\textsc{model\_l} + \textsc{model\_s} +\textsc{model\_c}+ \textsc{powerlaw} + \textsc{zgauss} $\times$ 13 + \textsc{constant} $\times$ \textsc{contamination} )
\newline

\noindent The first \textsc{constant} component is used to consider possible cross-calibration uncertainties between the different instruments. We fix the constant to one for {\it Chandra}, and allow it to be free for \textit{NuSTAR}. The second \textsc{constant} component is used to exclude the spectra of the contaminating sources in the {\it Chandra} observations, since they did not contribute to the emission. Therefore we fixed it to zero for the \textit{Chandra} spectrum, and to one for the \textit{NuSTAR} spectra. In order to constrain the contamination contribution at higher energies, the Compton scattering parameters are allowed to vary (the best-fitting model parameters are quoted in Table\,\ref{t:cont_fits} of Appendix\,\ref{ap:contamination}). The additional seven \textsc{zgauss} components with respect to \textsc{M2}, are the lines listed in Table \,\ref{t:FeK_complex}.

The top panel of Figure\,\ref{fig:overall_fit} shows the best fit model and residuals. We obtain that the cross-calibration constant between \textit{NuSTAR} and \textit{Chandra} is $0.85\pm0.03$, consistent with previous studies \citep[e.g.,][]{2010Nevalainen,2015Madsen}.  
Our model is able to well reproduce the overall curvature of the \textit{NuSTAR} spectrum. The best-fit values of the parameters obtained are shown in Table\,\ref{t:heg_fits}. They show some differences with respect to those obtained by fitting only the HEG spectrum. However, a Compton-thick reprocessor ($ N_{\rm H,d} = 1.01^{+0.03}_{-0.24} \: \rm cm^{-2} $) with a high covering factor ($0.55^{+0.01}_{-0.05}$) is still needed to reproduce the broadband X-ray spectrum of Circinus.  The ratio between the normalization of the scattered component and that of the continuum is $0.6 \pm 0.04 \%$, consistent with the values expected for heavily obscured AGN ($<1\%$; \citealp[e.g.,][]{2007Ueda,2021Gupta}).

The lower panel of Figure\,\ref{fig:overall_fit} illustrates the best fit model. It can be seen that the transmitted component (e.g., photons that do not interact with matter) is unconstrained, as the reflected component is dominant. The contamination component never dominates the spectrum, but it is important at all energies, in particular at low energies, in agreement with \citet{2014ApJ7...91...81A}. Taking into account the BLR, which also reprocesses and absorbs the X-ray photons, the effective equatorial column density becomes $ N_{\rm H} = 1.47^{+0.03}_{-0.24} \times 10^{25} \: \rm cm^{-2}$.

We also tried fitting the \textit{NuSTAR} spectrum without including the contamination in the fit (i.e., fitting the model \textsc{M1}  + \textsc{G}$_{\rm Fe}$). The best-fitting parameters and model are shown in Appendix\,\ref{ap:fit_nocont} (see Table\,\ref{t:nustar_nocont} and Figure\,\ref{fig:overall_fit_nocont}). We find a good fit ($\rm stat/d.o.f=1951/2410$), but we see some residuals at $E\gtrsim 25 \rm \: keV$, as the model slightly underpredicts the data. These results suggest that the contamination has a non-negligeable effect at high energies. It should be stressed that there could be some degeneracy between the parameters of the nuclear model and those of the off-nuclear component.

\section{Discussion}\label{sect:discussion}

In this paper we have built and tested, for the first time, an X-ray spectral model of an AGN starting from the structure of the circumnuclear material inferred from high-resolution observations carried out at lower energies. This was done for the Circinus galaxy, considering the recent models published by \citet{2017MNRAS.472.3854S,2019MNRAS.484.3334S} using interferometric and single-dish MIR observations, and combining those with our current understanding of the properties of the BLR and of the accretion disk.

Our spectral fitting of the {\it Chandra} and {\it NuSTAR} data shows that the MIR geometry is able to reproduce accurately most of the 2--70 keV X-ray spectra of Circinus. Several Gaussian lines between 6--7.1\,keV are however still needed to improve the fit around the Fe K$\alpha$ complex in the \textit{Chandra} HEG spectrum. Five of these lines (Fe IV-V, Fe VII-XII, FeXXV, Fe XXVI and Fe K$\beta$ II) are not reproduced by our model, since \textsc{RefleX} does not deal with ionized material, and they are attributed to ionized iron. Given the constraints of the FWHM we obtain, those lines likely originate from material associated with the outer regions of the BLR or torus atmosphere.  Figure\,\ref{fig:fe_eeuf} shows that to model the Fe K$\alpha$ and Fe K$\beta$ lines, additional Gaussian lines at 6.4 and 7.05 keV are needed. As discussed in Section\,\ref{sec:chandra_fit},  \citet{2014ApJ7...91...81A} finds that the overall broadening of the lines is due to their spatial extension along the dispersion direction, which is supported by the fact that the \kalfa{} line width of the 12-pixels aperture spectrum is higher than the line width of the 2-pixels aperture spectrum at $1-\sigma$ level. Also, the studies of \citet{2013Marinucci} and \citet{2021Uematsu} concluded that the Fe K$\alpha$ emission is extended by comparing the observed radial profile of Circinus with the \textit{Chandra} point spread function (PSF). A similar conclusion was drawn by Andonie et al. (submitted), who analyzed the  \textit{Chandra} radial profiles of Circinus and found that the \kalfa{} extends out to $\sim 100 \rm \: pc$.

An alternative explanation for the Fe K$\alpha$ line broadening may be that a portion of the line originates in the BLR, and is therefore broadened by the velocity of the clouds, which is not taken into account in our simulations. Previous works support the idea that a significant fraction of the narrow line may originate in the BLR or in the inner regions of the system \citep[e.g.][]{2010ApJS..187..581S,2015ApJ...812..113G}. We address this issue in Section\,\ref{sec:chandra_fit} by creating an additional table model for the BLR emission, and convolving it with the \textsc{gsmooth} model in XSPEC. The fit did not substantially improve, supporting the idea that the Fe K$\alpha$ line shape is affected by their spatial extension. Nonetheless, it should also be taken into account that the BLR does not have an uniform velocity, as it is assumed by \textsc{gsmooth}. To properly test this hypothesis, a model that considers different velocity fields is needed. We computed the FWHM of the Fe K$\alpha$ line and obtain a value of $v_{\rm FWHM} = 1491 \pm 87 \rm \: km\: s^{-1}$. Assuming virial motion, we can estimate the location of the emission radius of the line, $R =GM_{\rm BH}/(\sqrt{3}/2\:  v_{\rm FWHM})^2$ \citep[e.g.,][]{Netzer1990,2004ApJ...613..682P}. The SMBH mass of Circinus is $M_{\rm BH} = 10^{6.23}\:M_{\odot}$ \citep{2017ApJ...850...74K}, therefore, the emission radius of the Fe K$\alpha$ line is $R=(4.4 \pm 0.5)\times 10^{-3} \rm \: pc$, which corresponds to a region in the BLR (which has an inner radius of $R_{\rm BLR} = 3.8\times 10^{-3} \rm \: pc$). \citet{2011ApJ...738..147S} found that the bulk of the Fe K$\alpha$  typically arises from a region that is $\sim 0.7 - 11$ times the size of the optical BLR, which is consistent with our findings. Considering the very high inclination angle at which the system is observed (e.g., \citealp{2003Greenhill}), we exclude that reprocessing in the accretion disk might contribute significantly to the broadening of the Fe\,K$\alpha$ line (e.g., \citealp{Dauser:2010nt,Garcia:2014cf}). Moreover, the typical width of such broadened relativistic lines is $\gtrsim 10,000\rm km\,s^{-1}$ (e.g., \citealp{Fabian:1995ny,Fabian:2000hr,Nandra:1997md,Walton:2013qj}), with some object showing values up to $FWHM\sim 100,000\rm\,km\,s^{-1}$ (e.g., \citealp{Tanaka:1995wm,Ricci:2014cf,Walton:2019gv}), which is considerably higher than the value found for Circinus.

Our model is able to reproduce well the curvature of the spectra above 10\,keV. In the recent work of \citet{2019A&A...629A..16B} it has been shown that to model the Compton hump of Circinus and others local Compton-thick AGN, a Compton-thick reflector closer to the X-ray corona than the torus, with a relatively large covering factor ($>30\%$) is needed. Later, \citet{2021BuchnerB} developed a model considering a radiation pressure-driven polar outflow and found that the model only reproduces a good fit if a thick inner ring is included. We find that if such a reflector is not included, the model would underestimate the flux of the Compton hump, in agreement with what was shown by \citet{2019A&A...629A..16B,2021BuchnerB}. In our model, this component is associated with the accretion disk and the BLR. This suggests that future X-ray models of the circumnuclear environment of AGN should incorporate these components.  In a recent work, \citet{2021Uematsu} built an X-ray model with the aim of constraining the location of the torus of Circinus. They found that the torus is Compton-thick ($ N_{\rm H,equ} = 2.16^{+0.24}_{-0.17} \times 10^{25}\:  \rm cm^{-2}  $) and its angular width is $10.3^{+0.8}_{-0.3}$ degrees, which is much lower than the angular width we find. This difference could be due to the fact that the metallicity, which is strongly degenerate with $ N_{\rm H,equ}$ and $CF$, is a free parameter in the model of \citet{2021Uematsu}, who found a best-fit value of $Z=1.52 \: Z_{\odot}$. Additionally, they found that the Fe K$\alpha$ line originates in a region inside the dust sublimation radius, and concluded that the inner region of the torus is dust-free. In our model, that dust-free region is modelled by the BLR.

The final column density (considering both flared disk and BLR) is $ N_{\rm H} = 1.47^{+0.03}_{-0.24} \times 10^{25} \: \rm cm^{-2}$, which is consistent with Circinus being a Compton-thick AGN, as shown by numerous previous studies \citep[e.g.,][]{1999A&A...341L..39M,2000MNRAS.318..173M,2014ApJ7...91...81A,2021Uematsu,2021BuchnerB}, and the column density of the flared disk has the same order of magnitude as the value predicted by the small scale MIR model of S19. We obtain that the half-angular width of the flared disk is $33.3^{+1.4}_{-0.4}\rm\,degrees$ (equivalent to $CF=0.55^{+0.01}_{-0.05}$), which is higher than what was found in the IR by S17, but consistent with S19. In S19, the best fit value for the covering factor of the flared disk is $CF_{\rm S19} \sim 0.1$, but taking into account the homogeneous hyperboloid surface, the covering factor increase to $ CF_{\rm S19} \sim 0.87$. However, to account for the AGN near infrared emission, the hyperboloid must be clumpy, therefore, the total covering factor of the system (flared disk plus hyperboloid) is probably closer to the one inferred by our X-ray spectral modelling. \citet{2014ApJ7...91...81A} and \citet{2019A&A...629A..16B} were also able to reproduce the X-ray emission of Circinus with high covering factor reprocessors ($ CF\gtrsim 0.5-0.7$). The value we obtain for the radial column density of the cone is in between the values predicted by S17 and S19. We do not think this is a problem, as the MIR values are an starting point in our fits.

We find that the contamination (i.e., point sources and off-nuclear reflection) affecting the \textit{NuSTAR} observations never dominates the spectrum but it is important at all energies. Several previous studies aimed focussed on the \textit{NuSTAR} spectrum of Circinus did not include such a component in the fits, or only included the brightest point sources (i.e., the X-ray binary and supernova remnant). On the other hand, in our work, we perform a joint fit between the contamination and \textit{NuSTAR} spectra in order to separate the nuclear spectrum from the contaminating sources.

We find that the intrinsic 2--10\,keV luminosity of Circinus is $L_{2-10}=\rm (4.7^{+1}_{-0.7}) \times 10^{42}\:erg\:s^{-1}$, consistent with previous studies \citep[e.g.,][]{2014ApJ7...91...81A,2021Uematsu} and with the known observational MIR--X-ray correlation \citep[e.g.,][]{2009A&A...502..457G,2015MNRAS.454..766A}. In fact, using the relation found in \citet{2015MNRAS.454..766A}, from the 12$\mu$m luminosity of Circinus ($L_{12\mu \rm m} =(4.5 \pm 0.5)\times 10^{42}\,\rm erg\,s^{-1}$;  \citealp{2015MNRAS.454..766A}), one would expect an X-ray luminosity of $L_{2-10}=\rm (2.2_{-1.63}^{+4})\times 10^{42}\,\rm erg\,s^{-1}$.

\section{Summary and conclusions}\label{sec:conclusions}

We created a new model to reproduce the X-ray spectrum of the Circinus Galaxy based on the phenomenological infrared model released by \citet{2017MNRAS.472.3854S,2019MNRAS.484.3334S}, using the ray-tracing simulation platform \textsc{RefleX} \citep{2017A&A...607A..31P}. The model (see Figure\,\ref{fig:xr_model}) consists of three components in the equatorial plane: the accretion disk, the BLR, and a flared disk, the latter substituting the torus. We also included a hollow cone in the polar direction, consistent with recent observations of Circinus \citep{2007A&A...474..837T,2014A&A...563A..82T,2017MNRAS.472.3854S,2019MNRAS.484.3334S}. We applied this model to the broad-band X-ray spectrum of Circinus using data from \textit{Chandra} and \textit{NuSTAR} (see Section\,\ref{sec:xrayspectralfitting}), finding that:

\begin{enumerate}
    \item The model properly reproduces the curvature of the spectrum, the Fe K$\alpha$ line and its CS. The addition of Five Gaussian emission lines is however needed to reproduce the ionized iron lines, while two additional broad Gaussian emission lines at 6.4\,keV and 7.05\,keV (Fe K$\alpha$ and Fe K$\beta$ lines, $ v_{\rm FWHM}=1643\rm\,\:km \:s^{-1}$), are required to reproduce the Fe K$\alpha$ complex. We conclude that a possible spatial extension of the line along the HEG arms could play an important role on shape of the Fe K$\alpha$ line, and it needs to be taken into account to accurately reproduce it. Nevertheless, it remains unclear what fraction of this broadening might be caused by a spatial extension of the line, or due to Doppler broadening if part of the line originates in the BLR.
    \item We obtain a photon index of $\Gamma=1.65^{+0.11}_{-0.03}$, which is consistent with the median value observed in nearby obscured AGN \citep[e.g.,][]{2017ApJS..233...17R}.

    \item We find that the equatorial column density of Circinus is $ N_{\rm H} = 1.47^{+0.03}_{-0.24} \times 10^{25} \: \rm cm^{-2}$ and covering factor of the flared disk is $CF=0.55^{+0.01}_{-0.05}$, consistent with previous studies. 
    \item The best fitting values of the parameters of the flared disk (column density and covering factor) are close to the values obtained in the S19 modeling ($N_{\rm H,d,S19}= \rm 1.68\times10^{24}\:cm^{-2}$ and $ CF_{\rm S19} < 0.87$). Similarly, the best-fit value of the column density of cone ($ N_{\rm H,c}= \rm 2.18^{+0.47}_{-0.43}\times 10^{23}\: cm^{-2}$) is in between the values derived in S17 and S19 (i.e., $N_{\rm H,c,S17}= \rm 1.19\times10^{22}\: cm^{-2}$ and $ N_{\rm H,c,S19}= \rm 6.39\times10^{23}\:cm^{-2}$).
    
    \item We find an intrinsic X-ray luminosity of $L_{2-10}=\rm (4.7_{-0.7}^{+1}) \times 10^{42}\:erg\:s^{-1}$, consistent with what is expected from the $12 \: \mu \rm m$ luminosity of Circinus. 
    
    \item The contamination spectrum (including off-nuclear reflection and point sources) contributes significantly to the {\it NuSTAR} spectrum (see Figure\,\ref{fig:overall_fit}). Our broadband model that does not consider the contamination still provides an acceptable result, although it under-predicts the data at $E \gtrsim 25 \rm \: keV$ (see Figure\,\ref{fig:overall_fit_nocont}). Therefore, the contamination spectrum should always be considered, even when only the spectrum above $E \gtrsim 10$\,keV is being modeled.

\end{enumerate}

This work highlights the importance of building self-consistent physical models of AGN, in particular to analyze the high-resolution X-ray spectra that will be provided by future X-ray satellites such as {\it XRISM} and {\it Athena}/X-IFU \citep{2018SPIE10699E..1GB}. In particular, with calorimeters such as the one on board XRISM, it will be possible to fully constrain the different ionization levels of iron and their contribution to the 6.4-7\,keV flux, and also to understand what causes the shape of the Fe K$\alpha$ line and of its CS in Circinus: spatial extention of the line, gravitational broadening, or a combination of both. In an upcoming release of \textsc{RefleX} the effect of dust in the X-ray will also be included (Ricci \& Paltani in prep.), which will allow in the future to build even more sophisticated and realistic models of AGN.

\section*{Acknowledgements}

This project has received funding from the European Union’s Horizon 2020 research and innovation programme under the Marie Skłodowska-Curie grant agreement No 860744.

CR acknowledges support from the Fondecyt Iniciacion grant 11190831 and ANID BASAL project FB210003. The Geryon/Geryon2 cluster housed at the Centro de Astro-Ingenieria UC was used for (part) the calculations performed in this work. The BASAL PFB-06 CATA, Anillo ACT-86, FONDEQUIP AIC-57, and QUIMAL 130008 provided funding for several improvements to the Geryon/Geryon2 cluster.

This work was partially funded by
ANID - Millennium Science Initiative Program - ICN12\_009 (FEB), CATA-Basal - AFB-170002 (FEB, ET), and FONDECYT Regular - 1190818 (ET, FEB) and 1200495 (CA, FEB, ET).

ET acknowledge support from CATA-Basal AFB-170002, FONDECYT Regular grant 1190818, ANID Anillo ACT172033 and Millennium Nucleus NCN19\_058 (TITANs).

MS is supported by the Ministry of Education, Science and Technological Development of the Republic of Serbia through the contract no. 451-03-9/2021-14/200002 and the Science Fund of the Republic of Serbia, PROMIS 6060916,
BOWIE.

\section*{Data Availability}

The datasets generated and/or analysed in this study are available from the corresponding author on reasonable re- quest.

\bibliographystyle{mnras}
\bibliography{biblio}

\begin{thebibliography}{}
\makeatletter
\relax
\def\mn@urlcharsother{\let\do\@makeother \do\$\do\&\do\#\do\^\do\_\do\%\do\~}
\def\mn@doi{\begingroup\mn@urlcharsother \@ifnextchar [ {\mn@doi@}
  {\mn@doi@[]}}
\def\mn@doi@[#1]#2{\def\@tempa{#1}\ifx\@tempa\@empty \href
  {http://dx.doi.org/#2} {doi:#2}\else \href {http://dx.doi.org/#2} {#1}\fi
  \endgroup}
\def\mn@eprint#1#2{\mn@eprint@#1:#2::\@nil}
\def\mn@eprint@arXiv#1{\href {http://arxiv.org/abs/#1} {{\tt arXiv:#1}}}
\def\mn@eprint@dblp#1{\href {http://dblp.uni-trier.de/rec/bibtex/#1.xml}
  {dblp:#1}}
\def\mn@eprint@#1:#2:#3:#4\@nil{\def\@tempa {#1}\def\@tempb {#2}\def\@tempc
  {#3}\ifx \@tempc \@empty \let \@tempc \@tempb \let \@tempb \@tempa \fi \ifx
  \@tempb \@empty \def\@tempb {arXiv}\fi \@ifundefined
  {mn@eprint@\@tempb}{\@tempb:\@tempc}{\expandafter \expandafter \csname
  mn@eprint@\@tempb\endcsname \expandafter{\@tempc}}}

\bibitem[\protect\citeauthoryear{{Antonucci}}{{Antonucci}}{1993}]{Antonucci:1993fu}
{Antonucci} R.,  1993, \mn@doi [\araa] {10.1146/annurev.aa.31.090193.002353},
  \href {http://adsabs.harvard.edu/abs/1993ARA%26A..31..473A} {31, 473}

\bibitem[\protect\citeauthoryear{{Ar{\'e}valo} et~al.,}{{Ar{\'e}valo}
  et~al.}{2014}]{2014ApJ7...91...81A}
{Ar{\'e}valo} P.,  et~al., 2014, \mn@doi [\apj] {10.1088/0004-637X/791/2/81},
  \href {http://adsabs.harvard.edu/abs/2014ApJ...791...81A} {791, 81}

\bibitem[\protect\citeauthoryear{{Arnaud}}{{Arnaud}}{1996}]{1996ASPC..101...17A}
{Arnaud} K.~A.,  1996, in {Jacoby} G.~H.,  {Barnes} J.,  eds,  Astronomical
  Society of the Pacific Conference Series Vol. 101, Astronomical Data Analysis
  Software and Systems V. p.~17

\bibitem[\protect\citeauthoryear{{Asmus}}{{Asmus}}{2019}]{2019MNRAS.489.2177A}
{Asmus} D.,  2019, \mn@doi [\mnras] {10.1093/mnras/stz2289}, \href
  {https://ui.adsabs.harvard.edu/abs/2019MNRAS.489.2177A} {489, 2177}

\bibitem[\protect\citeauthoryear{{Asmus}, {Gandhi}, {H{\"o}nig}, {Smette}  \&
  {Duschl}}{{Asmus} et~al.}{2015}]{2015MNRAS.454..766A}
{Asmus} D.,  {Gandhi} P.,  {H{\"o}nig} S.~F.,  {Smette} A.,   {Duschl} W.~J.,
  2015, \mn@doi [\mnras] {10.1093/mnras/stv1950}, \href
  {http://adsabs.harvard.edu/abs/2015MNRAS.454..766A} {454, 766}

\bibitem[\protect\citeauthoryear{{Asmus}, {H{\"o}nig}  \& {Gandhi}}{{Asmus}
  et~al.}{2016}]{2016ApJ...822..109A}
{Asmus} D.,  {H{\"o}nig} S.~F.,   {Gandhi} P.,  2016, \mn@doi [\apj]
  {10.3847/0004-637X/822/2/109}, \href
  {https://ui.adsabs.harvard.edu/abs/2016ApJ...822..109A} {822, 109}

\bibitem[\protect\citeauthoryear{{Awaki}, {Kunieda}, {Tawara}  \&
  {Koyama}}{{Awaki} et~al.}{1991}]{1991PASJ...43L..37A}
{Awaki} H.,  {Kunieda} H.,  {Tawara} Y.,   {Koyama} K.,  1991, \pasj, \href
  {https://ui.adsabs.harvard.edu/abs/1991PASJ...43L..37A} {43, L37}

\bibitem[\protect\citeauthoryear{{Balokovi{\'c}} et~al.,}{{Balokovi{\'c}}
  et~al.}{2018}]{2018ApJ...854...42B}
{Balokovi{\'c}} M.,  et~al., 2018, \mn@doi [\apj] {10.3847/1538-4357/aaa7eb},
  \href {https://ui.adsabs.harvard.edu/abs/2018ApJ...854...42B} {854, 42}

\bibitem[\protect\citeauthoryear{{Barret} et~al.,}{{Barret}
  et~al.}{2018}]{2018SPIE10699E..1GB}
{Barret} D.,  et~al., 2018, in {den Herder} J.-W.~A.,  {Nikzad} S.,
  {Nakazawa} K.,  eds,  Society of Photo-Optical Instrumentation Engineers
  (SPIE) Conference Series Vol. 10699, Space Telescopes and Instrumentation
  2018: Ultraviolet to Gamma Ray. p. 106991G (\mn@eprint {arXiv} {1807.06092}),
  \mn@doi{10.1117/12.2312409}

\bibitem[\protect\citeauthoryear{{Bauer}, {Brandt}, {Sambruna}, {Chartas},
  {Garmire}, {Kaspi}  \& {Netzer}}{{Bauer} et~al.}{2001}]{2001AJ....122..182B}
{Bauer} F.~E.,  {Brandt} W.~N.,  {Sambruna} R.~M.,  {Chartas} G.,  {Garmire}
  G.~P.,  {Kaspi} S.,   {Netzer} H.,  2001, \mn@doi [\aj] {10.1086/321123},
  \href {http://adsabs.harvard.edu/abs/2001AJ....122..182B} {122, 182}

\bibitem[\protect\citeauthoryear{{Bauer} et~al.,}{{Bauer}
  et~al.}{2015}]{2015Bauer}
{Bauer} F.~E.,  et~al., 2015, \mn@doi [\apj] {10.1088/0004-637X/812/2/116},
  \href {https://ui.adsabs.harvard.edu/abs/2015ApJ...812..116B} {812, 116}

\bibitem[\protect\citeauthoryear{{Brightman} \& {Nandra}}{{Brightman} \&
  {Nandra}}{2011}]{2011MNRAS.413.1206B}
{Brightman} M.,  {Nandra} K.,  2011, \mn@doi [\mnras]
  {10.1111/j.1365-2966.2011.18207.x}, \href
  {http://adsabs.harvard.edu/abs/2011MNRAS.413.1206B} {413, 1206}

\bibitem[\protect\citeauthoryear{{Buchner}, {Brightman}, {Nandra}, {Nikutta}
  \& {Bauer}}{{Buchner} et~al.}{2019}]{2019A&A...629A..16B}
{Buchner} J.,  {Brightman} M.,  {Nandra} K.,  {Nikutta} R.,   {Bauer} F.~E.,
  2019, \mn@doi [\aap] {10.1051/0004-6361/201834771}, \href
  {https://ui.adsabs.harvard.edu/abs/2019A&A...629A..16B} {629, A16}

\bibitem[\protect\citeauthoryear{{Buchner}, {Brightman}, {Balokovi{\'c}},
  {Wada}, {Bauer}  \& {Nandra}}{{Buchner} et~al.}{2021}]{2021BuchnerB}
{Buchner} J.,  {Brightman} M.,  {Balokovi{\'c}} M.,  {Wada} K.,  {Bauer} F.~E.,
    {Nandra} K.,  2021, arXiv e-prints, \href
  {https://ui.adsabs.harvard.edu/abs/2021arXiv210608331B} {p. arXiv:2106.08331}

\bibitem[\protect\citeauthoryear{{Burtscher} et~al.,}{{Burtscher}
  et~al.}{2013}]{2013A&A...558A.149B}
{Burtscher} L.,  et~al., 2013, \mn@doi [\aap] {10.1051/0004-6361/201321890},
  \href {https://ui.adsabs.harvard.edu/abs/2013A&A...558A.149B} {558, A149}

\bibitem[\protect\citeauthoryear{{Cash}}{{Cash}}{1979}]{1979Cash}
{Cash} W.,  1979, \mn@doi [\apj] {10.1086/156922}, \href
  {http://adsabs.harvard.edu/abs/1979ApJ...228..939C} {228, 939}

\bibitem[\protect\citeauthoryear{{Chartas}, {Kochanek}, {Dai}, {Poindexter}  \&
  {Garmire}}{{Chartas} et~al.}{2009}]{2009ApJ...693..174C}
{Chartas} G.,  {Kochanek} C.~S.,  {Dai} X.,  {Poindexter} S.,   {Garmire} G.,
  2009, \mn@doi [\apj] {10.1088/0004-637X/693/1/174}, \href
  {http://adsabs.harvard.edu/abs/2009ApJ...693..174C} {693, 174}

\bibitem[\protect\citeauthoryear{{Dauser}, {Wilms}, {Reynolds}  \&
  {Brenneman}}{{Dauser} et~al.}{2010}]{Dauser:2010nt}
{Dauser} T.,  {Wilms} J.,  {Reynolds} C.~S.,   {Brenneman} L.~W.,  2010,
  \mn@doi [\mnras] {10.1111/j.1365-2966.2010.17393.x}, \href
  {https://ui.adsabs.harvard.edu/abs/2010MNRAS.409.1534D} {409, 1534}

\bibitem[\protect\citeauthoryear{{De Marco}, {Ponti}, {Cappi}, {Dadina},
  {Uttley}, {Cackett}, {Fabian}  \& {Miniutti}}{{De Marco}
  et~al.}{2013}]{2013MNRAS.431.2441D}
{De Marco} B.,  {Ponti} G.,  {Cappi} M.,  {Dadina} M.,  {Uttley} P.,  {Cackett}
  E.~M.,  {Fabian} A.~C.,   {Miniutti} G.,  2013, \mn@doi [\mnras]
  {10.1093/mnras/stt339}, \href
  {http://adsabs.harvard.edu/abs/2013MNRAS.431.2441D} {431, 2441}

\bibitem[\protect\citeauthoryear{{Dunn}, {Crenshaw}, {Kraemer}  \&
  {Gabel}}{{Dunn} et~al.}{2007}]{2007AJ....134.1061D}
{Dunn} J.~P.,  {Crenshaw} D.~M.,  {Kraemer} S.~B.,   {Gabel} J.~R.,  2007,
  \mn@doi [\aj] {10.1086/520644}, \href
  {https://ui.adsabs.harvard.edu/abs/2007AJ....134.1061D} {134, 1061}

\bibitem[\protect\citeauthoryear{{Esparza-Arredondo}
  et~al.,}{{Esparza-Arredondo} et~al.}{2019}]{Esparza-Arredondo:2019ux}
{Esparza-Arredondo} D.,  et~al., 2019, \mn@doi [\apj]
  {10.3847/1538-4357/ab4ced}, \href
  {https://ui.adsabs.harvard.edu/abs/2019ApJ...886..125E} {886, 125}

\bibitem[\protect\citeauthoryear{{Esparza-Arredondo}, {Gonzalez-Mart{\'\i}n},
  {Dultzin}, {Masegosa}, {Ramos-Almeida}, {Garc{\'\i}a-Bernete}, {Fritz}  \&
  {Osorio-Clavijo}}{{Esparza-Arredondo}
  et~al.}{2021}]{Esparza-Arredondo:2021ep}
{Esparza-Arredondo} D.,  {Gonzalez-Mart{\'\i}n} O.,  {Dultzin} D.,  {Masegosa}
  J.,  {Ramos-Almeida} C.,  {Garc{\'\i}a-Bernete} I.,  {Fritz} J.,
  {Osorio-Clavijo} N.,  2021, \mn@doi [\aap] {10.1051/0004-6361/202040043},
  \href {https://ui.adsabs.harvard.edu/abs/2021A&A...651A..91E} {651, A91}

\bibitem[\protect\citeauthoryear{{Fabbiano}, {Paggi}, {Karovska}, {Elvis},
  {Maksym}, {Risaliti}  \& {Wang}}{{Fabbiano} et~al.}{2018a}]{2018BFabbiano}
{Fabbiano} G.,  {Paggi} A.,  {Karovska} M.,  {Elvis} M.,  {Maksym} W.~P.,
  {Risaliti} G.,   {Wang} J.,  2018a, \mn@doi [\apj]
  {10.3847/1538-4357/aab1f4}, \href
  {https://ui.adsabs.harvard.edu/abs/2018ApJ...855..131F} {855, 131}

\bibitem[\protect\citeauthoryear{{Fabbiano}, {Paggi}, {Karovska}, {Elvis},
  {Maksym}  \& {Wang}}{{Fabbiano} et~al.}{2018b}]{2018Fabbiano}
{Fabbiano} G.,  {Paggi} A.,  {Karovska} M.,  {Elvis} M.,  {Maksym} W.~P.,
  {Wang} J.,  2018b, \mn@doi [\apj] {10.3847/1538-4357/aadc5d}, \href
  {https://ui.adsabs.harvard.edu/abs/2018ApJ...865...83F} {865, 83}

\bibitem[\protect\citeauthoryear{{Fabian}, {Nandra}, {Reynolds}, {Brandt},
  {Otani}, {Tanaka}, {Inoue}  \& {Iwasawa}}{{Fabian}
  et~al.}{1995}]{Fabian:1995ny}
{Fabian} A.~C.,  {Nandra} K.,  {Reynolds} C.~S.,  {Brandt} W.~N.,  {Otani} C.,
  {Tanaka} Y.,  {Inoue} H.,   {Iwasawa} K.,  1995, \mn@doi [\mnras]
  {10.1093/mnras/277.1.L11}, \href
  {https://ui.adsabs.harvard.edu/abs/1995MNRAS.277L..11F} {277, L11}

\bibitem[\protect\citeauthoryear{{Fabian}, {Iwasawa}, {Reynolds}  \&
  {Young}}{{Fabian} et~al.}{2000}]{Fabian:2000hr}
{Fabian} A.~C.,  {Iwasawa} K.,  {Reynolds} C.~S.,   {Young} A.~J.,  2000,
  \mn@doi [\pasp] {10.1086/316610}, \href
  {https://ui.adsabs.harvard.edu/abs/2000PASP..112.1145F} {112, 1145}

\bibitem[\protect\citeauthoryear{{Fabian}, {Celotti}  \& {Erlund}}{{Fabian}
  et~al.}{2006}]{2006MNRAS.373L..16F}
{Fabian} A.~C.,  {Celotti} A.,   {Erlund} M.~C.,  2006, \mn@doi [\mnras]
  {10.1111/j.1745-3933.2006.00234.x}, \href
  {https://ui.adsabs.harvard.edu/abs/2006MNRAS.373L..16F} {373, L16}

\bibitem[\protect\citeauthoryear{{Fabian}, {Vasudevan}  \& {Gandhi}}{{Fabian}
  et~al.}{2008}]{2008MNRAS.385L..43F}
{Fabian} A.~C.,  {Vasudevan} R.~V.,   {Gandhi} P.,  2008, \mn@doi [\mnras]
  {10.1111/j.1745-3933.2008.00430.x}, \href
  {https://ui.adsabs.harvard.edu/abs/2008MNRAS.385L..43F} {385, L43}

\bibitem[\protect\citeauthoryear{{Fabian} et~al.,}{{Fabian}
  et~al.}{2009}]{2009Natur.459..540F}
{Fabian} A.~C.,  et~al., 2009, \mn@doi [\nat] {10.1038/nature08007}, \href
  {http://adsabs.harvard.edu/abs/2009Natur.459..540F} {459, 540}

\bibitem[\protect\citeauthoryear{{Farrah} et~al.,}{{Farrah}
  et~al.}{2016}]{Farrah:2016nw}
{Farrah} D.,  et~al., 2016, \mn@doi [\apj] {10.3847/0004-637X/831/1/76}, \href
  {https://ui.adsabs.harvard.edu/abs/2016ApJ...831...76F} {831, 76}

\bibitem[\protect\citeauthoryear{{Ferland}, {Peterson}, {Horne}, {Welsh}  \&
  {Nahar}}{{Ferland} et~al.}{1992}]{1992ApJ...387...95F}
{Ferland} G.~J.,  {Peterson} B.~M.,  {Horne} K.,  {Welsh} W.~F.,   {Nahar}
  S.~N.,  1992, \mn@doi [\apj] {10.1086/171063}, \href
  {http://adsabs.harvard.edu/abs/1992ApJ...387...95F} {387, 95}

\bibitem[\protect\citeauthoryear{{Freeman}, {Karlsson}, {Lynga}, {Burrell},
  {van Woerden}, {Goss}  \& {Mebold}}{{Freeman}
  et~al.}{1977}]{1977A&A....55..445F}
{Freeman} K.~C.,  {Karlsson} B.,  {Lynga} G.,  {Burrell} J.~F.,  {van Woerden}
  H.,  {Goss} W.~M.,   {Mebold} U.,  1977, \aap, \href
  {https://ui.adsabs.harvard.edu/abs/1977A&A....55..445F} {55, 445}

\bibitem[\protect\citeauthoryear{{Gandhi}, {Horst}, {Smette}, {H{\"o}nig},
  {Comastri}, {Gilli}, {Vignali}  \& {Duschl}}{{Gandhi}
  et~al.}{2009}]{2009A&A...502..457G}
{Gandhi} P.,  {Horst} H.,  {Smette} A.,  {H{\"o}nig} S.,  {Comastri} A.,
  {Gilli} R.,  {Vignali} C.,   {Duschl} W.,  2009, \mn@doi [\aap]
  {10.1051/0004-6361/200811368}, \href
  {http://adsabs.harvard.edu/abs/2009A%26A...502..457G} {502, 457}

\bibitem[\protect\citeauthoryear{{Gandhi}, {H{\"o}nig}  \&
  {Kishimoto}}{{Gandhi} et~al.}{2015}]{2015ApJ...812..113G}
{Gandhi} P.,  {H{\"o}nig} S.~F.,   {Kishimoto} M.,  2015, \mn@doi [\apj]
  {10.1088/0004-637X/812/2/113}, \href
  {https://ui.adsabs.harvard.edu/abs/2015ApJ...812..113G} {812, 113}

\bibitem[\protect\citeauthoryear{{Garc{\'{\i}}a} \& {Kallman}}{{Garc{\'{\i}}a}
  \& {Kallman}}{2010}]{2010ApJ...718..695G}
{Garc{\'{\i}}a} J.,  {Kallman} T.~R.,  2010, \mn@doi [\apj]
  {10.1088/0004-637X/718/2/695}, \href
  {http://adsabs.harvard.edu/abs/2010ApJ...718..695G} {718, 695}

\bibitem[\protect\citeauthoryear{{Garc{\'\i}a} et~al.,}{{Garc{\'\i}a}
  et~al.}{2014}]{Garcia:2014cf}
{Garc{\'\i}a} J.,  et~al., 2014, \mn@doi [\apj] {10.1088/0004-637X/782/2/76},
  \href {https://ui.adsabs.harvard.edu/abs/2014ApJ...782...76G} {782, 76}

\bibitem[\protect\citeauthoryear{{George} \& {Fabian}}{{George} \&
  {Fabian}}{1991}]{1991MNRAS.249..352G}
{George} I.~M.,  {Fabian} A.~C.,  1991, \mn@doi [\mnras]
  {10.1093/mnras/249.2.352}, \href
  {https://ui.adsabs.harvard.edu/abs/1991MNRAS.249..352G} {249, 352}

\bibitem[\protect\citeauthoryear{{Ghisellini}, {Haardt}  \&
  {Matt}}{{Ghisellini} et~al.}{1994}]{1994MNRAS.267..743G}
{Ghisellini} G.,  {Haardt} F.,   {Matt} G.,  1994, \mn@doi [\mnras]
  {10.1093/mnras/267.3.743}, \href
  {https://ui.adsabs.harvard.edu/abs/1994MNRAS.267..743G} {267, 743}

\bibitem[\protect\citeauthoryear{{Goad}, {Korista}  \& {Ruff}}{{Goad}
  et~al.}{2012}]{2012MNRAS.426.3086G}
{Goad} M.~R.,  {Korista} K.~T.,   {Ruff} A.~J.,  2012, \mn@doi [\mnras]
  {10.1111/j.1365-2966.2012.21808.x}, \href
  {http://adsabs.harvard.edu/abs/2012MNRAS.426.3086G} {426, 3086}

\bibitem[\protect\citeauthoryear{{Gravity Collaboration} et~al.,}{{Gravity
  Collaboration} et~al.}{2018}]{Gravity-Collaboration:2018ci}
{Gravity Collaboration} et~al., 2018, \mn@doi [\nat]
  {10.1038/s41586-018-0731-9}, \href
  {https://ui.adsabs.harvard.edu/abs/2018Natur.563..657G} {563, 657}

\bibitem[\protect\citeauthoryear{{Gravity Collaboration} et~al.,}{{Gravity
  Collaboration} et~al.}{2020}]{Gravity-Collaboration:2020dh}
{Gravity Collaboration} et~al., 2020, \mn@doi [\aap]
  {10.1051/0004-6361/202039067}, \href
  {https://ui.adsabs.harvard.edu/abs/2020A&A...643A.154G} {643, A154}

\bibitem[\protect\citeauthoryear{{Gravity Collaboration} et~al.,}{{Gravity
  Collaboration} et~al.}{2021}]{Gravity-Collaboration:2021pt}
{Gravity Collaboration} et~al., 2021, \mn@doi [\aap]
  {10.1051/0004-6361/202040061}, \href
  {https://ui.adsabs.harvard.edu/abs/2021A&A...648A.117G} {648, A117}

\bibitem[\protect\citeauthoryear{{Greenhill} et~al.,}{{Greenhill}
  et~al.}{2003}]{2003Greenhill}
{Greenhill} L.~J.,  et~al., 2003, \mn@doi [\apj] {10.1086/374862}, \href
  {https://ui.adsabs.harvard.edu/abs/2003ApJ...590..162G} {590, 162}

\bibitem[\protect\citeauthoryear{{Guainazzi} \& {Bianchi}}{{Guainazzi} \&
  {Bianchi}}{2007}]{2007MNRAS.374.1290G}
{Guainazzi} M.,  {Bianchi} S.,  2007, \mn@doi [\mnras]
  {10.1111/j.1365-2966.2006.11229.x}, \href
  {https://ui.adsabs.harvard.edu/abs/2007MNRAS.374.1290G} {374, 1290}

\bibitem[\protect\citeauthoryear{{Guainazzi}, {Matt}  \& {Perola}}{{Guainazzi}
  et~al.}{2005}]{2005A&A...444..119G}
{Guainazzi} M.,  {Matt} G.,   {Perola} G.~C.,  2005, \mn@doi [\aap]
  {10.1051/0004-6361:20053643}, \href
  {https://ui.adsabs.harvard.edu/abs/2005A&A...444..119G} {444, 119}

\bibitem[\protect\citeauthoryear{{Gupta} et~al.,}{{Gupta}
  et~al.}{2021}]{2021Gupta}
{Gupta} K.~K.,  et~al., 2021, \mn@doi [\mnras] {10.1093/mnras/stab839}, \href
  {https://ui.adsabs.harvard.edu/abs/2021MNRAS.tmp..839G} {}

\bibitem[\protect\citeauthoryear{{Haardt} \& {Maraschi}}{{Haardt} \&
  {Maraschi}}{1991}]{1991ApJ...380L..51H}
{Haardt} F.,  {Maraschi} L.,  1991, \mn@doi [\apjl] {10.1086/186171}, \href
  {https://ui.adsabs.harvard.edu/abs/1991ApJ...380L..51H} {380, L51}

\bibitem[\protect\citeauthoryear{{Hagiwara}, {Miyoshi}, {Doi}  \&
  {Horiuchi}}{{Hagiwara} et~al.}{2013}]{2013ApJ...768L..38H}
{Hagiwara} Y.,  {Miyoshi} M.,  {Doi} A.,   {Horiuchi} S.,  2013, \mn@doi
  [\apjl] {10.1088/2041-8205/768/2/L38}, \href
  {https://ui.adsabs.harvard.edu/abs/2013ApJ...768L..38H} {768, L38}

\bibitem[\protect\citeauthoryear{{Harrison} et~al.,}{{Harrison}
  et~al.}{2013}]{2013ApJ...770..103H}
{Harrison} F.~A.,  et~al., 2013, \mn@doi [\apj] {10.1088/0004-637X/770/2/103},
  \href {http://adsabs.harvard.edu/abs/2013ApJ...770..103H} {770, 103}

\bibitem[\protect\citeauthoryear{{H{\"o}nig} \& {Beckert}}{{H{\"o}nig} \&
  {Beckert}}{2007}]{2007MNRAS.380.1172H}
{H{\"o}nig} S.~F.,  {Beckert} T.,  2007, \mn@doi [\mnras]
  {10.1111/j.1365-2966.2007.12157.x}, \href
  {https://ui.adsabs.harvard.edu/abs/2007MNRAS.380.1172H} {380, 1172}

\bibitem[\protect\citeauthoryear{{H{\"o}nig}, {Kishimoto}, {Antonucci},
  {Marconi}, {Prieto}, {Tristram}  \& {Weigelt}}{{H{\"o}nig}
  et~al.}{2012}]{2012ApJ...755..149H}
{H{\"o}nig} S.~F.,  {Kishimoto} M.,  {Antonucci} R.,  {Marconi} A.,  {Prieto}
  M.~A.,  {Tristram} K.,   {Weigelt} G.,  2012, \mn@doi [\apj]
  {10.1088/0004-637X/755/2/149}, \href
  {https://ui.adsabs.harvard.edu/abs/2012ApJ...755..149H} {755, 149}

\bibitem[\protect\citeauthoryear{{Hutsem{\'e}kers} \&
  {Sluse}}{{Hutsem{\'e}kers} \& {Sluse}}{2021}]{Hutsemekers:2021do}
{Hutsem{\'e}kers} D.,  {Sluse} D.,  2021, \mn@doi [\aap]
  {10.1051/0004-6361/202141820}, \href
  {https://ui.adsabs.harvard.edu/abs/2021A&A...654A.155H} {654, A155}

\bibitem[\protect\citeauthoryear{{Jaffe} et~al.,}{{Jaffe}
  et~al.}{2004}]{2004Natur.429...47J}
{Jaffe} W.,  et~al., 2004, \mn@doi [\nat] {10.1038/nature02531}, \href
  {https://ui.adsabs.harvard.edu/abs/2004Natur.429...47J} {429, 47}

\bibitem[\protect\citeauthoryear{{Johnson} \& {Soff}}{{Johnson} \&
  {Soff}}{1985}]{1985Johnson}
{Johnson} W.~R.,  {Soff} G.,  1985, \mn@doi [Atomic Data and Nuclear Data
  Tables] {10.1016/0092-640X(85)90010-5}, \href
  {https://ui.adsabs.harvard.edu/abs/1985ADNDT..33..405J} {33, 405}

\bibitem[\protect\citeauthoryear{{Kaastra} \& {Mewe}}{{Kaastra} \&
  {Mewe}}{1993}]{1993Kaastra}
{Kaastra} J.~S.,  {Mewe} R.,  1993, \aaps, \href
  {https://ui.adsabs.harvard.edu/abs/1993A&AS...97..443K} {97, 443}

\bibitem[\protect\citeauthoryear{{Kalberla}, {Burton}, {Hartmann}, {Arnal},
  {Bajaja}, {Morras}  \& {P{\"o}ppel}}{{Kalberla}
  et~al.}{2005}]{2005A&A...440..775K}
{Kalberla} P.~M.~W.,  {Burton} W.~B.,  {Hartmann} D.,  {Arnal} E.~M.,  {Bajaja}
  E.,  {Morras} R.,   {P{\"o}ppel} W.~G.~L.,  2005, \mn@doi [\aap]
  {10.1051/0004-6361:20041864}, \href
  {https://ui.adsabs.harvard.edu/abs/2005A&A...440..775K} {440, 775}

\bibitem[\protect\citeauthoryear{{Kaspi}, {Maoz}, {Netzer}, {Peterson},
  {Vestergaard}  \& {Jannuzi}}{{Kaspi} et~al.}{2005}]{2005ApJ...629...61K}
{Kaspi} S.,  {Maoz} D.,  {Netzer} H.,  {Peterson} B.~M.,  {Vestergaard} M.,
  {Jannuzi} B.~T.,  2005, \mn@doi [\apj] {10.1086/431275}, \href
  {http://adsabs.harvard.edu/abs/2005ApJ...629...61K} {629, 61}

\bibitem[\protect\citeauthoryear{{Kawamuro}, {Izumi}  \& {Imanishi}}{{Kawamuro}
  et~al.}{2019}]{2019Kawamuro}
{Kawamuro} T.,  {Izumi} T.,   {Imanishi} M.,  2019, \mn@doi [\pasj]
  {10.1093/pasj/psz045}, \href
  {https://ui.adsabs.harvard.edu/abs/2019PASJ...71...68K} {71, 68}

\bibitem[\protect\citeauthoryear{{Koribalski} et~al.,}{{Koribalski}
  et~al.}{2004}]{2004AJ....128...16K}
{Koribalski} B.~S.,  et~al., 2004, \mn@doi [\aj] {10.1086/421744}, \href
  {https://ui.adsabs.harvard.edu/abs/2004AJ....128...16K} {128, 16}

\bibitem[\protect\citeauthoryear{{Kormendy} \& {Ho}}{{Kormendy} \&
  {Ho}}{2013}]{2013Kormendy}
{Kormendy} J.,  {Ho} L.~C.,  2013, \mn@doi [\araa]
  {10.1146/annurev-astro-082708-101811}, \href
  {https://ui.adsabs.harvard.edu/abs/2013ARA&A..51..511K} {51, 511}

\bibitem[\protect\citeauthoryear{{Koss} et~al.,}{{Koss}
  et~al.}{2017}]{2017ApJ...850...74K}
{Koss} M.,  et~al., 2017, \mn@doi [\apj] {10.3847/1538-4357/aa8ec9}, \href
  {https://ui.adsabs.harvard.edu/abs/2017ApJ...850...74K} {850, 74}

\bibitem[\protect\citeauthoryear{{Lagage} et~al.,}{{Lagage}
  et~al.}{2004}]{2004Msngr.117...12L}
{Lagage} P.~O.,  et~al., 2004, The Messenger, \href
  {https://ui.adsabs.harvard.edu/abs/2004Msngr.117...12L} {117, 12}

\bibitem[\protect\citeauthoryear{{Lanz} et~al.,}{{Lanz}
  et~al.}{2019}]{Lanz:2019ey}
{Lanz} L.,  et~al., 2019, \mn@doi [\apj] {10.3847/1538-4357/aaee6c}, \href
  {https://ui.adsabs.harvard.edu/abs/2019ApJ...870...26L} {870, 26}

\bibitem[\protect\citeauthoryear{{Lawther}, {Goad}, {Korista}, {Ulrich}  \&
  {Vestergaard}}{{Lawther} et~al.}{2018}]{2018MNRAS.481..533L}
{Lawther} D.,  {Goad} M.~R.,  {Korista} K.~T.,  {Ulrich} O.,   {Vestergaard}
  M.,  2018, \mn@doi [\mnras] {10.1093/mnras/sty2242}, \href
  {https://ui.adsabs.harvard.edu/abs/2018MNRAS.481..533L} {481, 533}

\bibitem[\protect\citeauthoryear{{Leftley}, {H{\"o}nig}, {Asmus}, {Tristram},
  {Gandhi}, {Kishimoto}, {Venanzi}  \& {Williamson}}{{Leftley}
  et~al.}{2019}]{2019ApJ...886...55L}
{Leftley} J.~H.,  {H{\"o}nig} S.~F.,  {Asmus} D.,  {Tristram} K. R.~W.,
  {Gandhi} P.,  {Kishimoto} M.,  {Venanzi} M.,   {Williamson} D.~J.,  2019,
  \mn@doi [\apj] {10.3847/1538-4357/ab4a0b}, \href
  {https://ui.adsabs.harvard.edu/abs/2019ApJ...886...55L} {886, 55}

\bibitem[\protect\citeauthoryear{{Liu} \& {Li}}{{Liu} \&
  {Li}}{2014}]{2014ApJ...787...52L}
{Liu} Y.,  {Li} X.,  2014, \mn@doi [\apj] {10.1088/0004-637X/787/1/52}, \href
  {https://ui.adsabs.harvard.edu/abs/2014ApJ...787...52L} {787, 52}

\bibitem[\protect\citeauthoryear{{Liu}, {H{\"o}nig}, {Ricci}  \&
  {Paltani}}{{Liu} et~al.}{2019}]{2019MNRAS.490.4344L}
{Liu} J.,  {H{\"o}nig} S.~F.,  {Ricci} C.,   {Paltani} S.,  2019, \mn@doi
  [\mnras] {10.1093/mnras/stz2908}, \href
  {https://ui.adsabs.harvard.edu/abs/2019MNRAS.490.4344L} {490, 4344}

\bibitem[\protect\citeauthoryear{{Lodders}}{{Lodders}}{2003}]{2003ApJ...591.1220L}
{Lodders} K.,  2003, \mn@doi [\apj] {10.1086/375492}, \href
  {https://ui.adsabs.harvard.edu/abs/2003ApJ...591.1220L} {591, 1220}

\bibitem[\protect\citeauthoryear{{L{\'o}pez-Gonzaga}, {Burtscher}, {Tristram},
  {Meisenheimer}  \& {Schartmann}}{{L{\'o}pez-Gonzaga}
  et~al.}{2016}]{2016A&A...591A..47L}
{L{\'o}pez-Gonzaga} N.,  {Burtscher} L.,  {Tristram} K.~R.~W.,  {Meisenheimer}
  K.,   {Schartmann} M.,  2016, \mn@doi [\aap] {10.1051/0004-6361/201527590},
  \href {https://ui.adsabs.harvard.edu/abs/2016A&A...591A..47L} {591, A47}

\bibitem[\protect\citeauthoryear{{Madsen} et~al.,}{{Madsen}
  et~al.}{2015}]{2015Madsen}
{Madsen} K.~K.,  et~al., 2015, \mn@doi [\apjs] {10.1088/0067-0049/220/1/8},
  \href {https://ui.adsabs.harvard.edu/abs/2015ApJS..220....8M} {220, 8}

\bibitem[\protect\citeauthoryear{{Magorrian} et~al.,}{{Magorrian}
  et~al.}{1998}]{1998AJ....115.2285M}
{Magorrian} J.,  et~al., 1998, \mn@doi [\aj] {10.1086/300353}, \href
  {https://ui.adsabs.harvard.edu/abs/1998AJ....115.2285M} {115, 2285}

\bibitem[\protect\citeauthoryear{{Maiolino}, {Marconi}, {Salvati}, {Risaliti},
  {Severgnini}, {Oliva}, {La Franca}  \& {Vanzi}}{{Maiolino}
  et~al.}{2001a}]{2001A&A...365...28M}
{Maiolino} R.,  {Marconi} A.,  {Salvati} M.,  {Risaliti} G.,  {Severgnini} P.,
  {Oliva} E.,  {La Franca} F.,   {Vanzi} L.,  2001a, \mn@doi [\aap]
  {10.1051/0004-6361:20000177}, \href
  {https://ui.adsabs.harvard.edu/abs/2001A&A...365...28M} {365, 28}

\bibitem[\protect\citeauthoryear{{Maiolino}, {Marconi}  \& {Oliva}}{{Maiolino}
  et~al.}{2001b}]{2001A&A...365...37M}
{Maiolino} R.,  {Marconi} A.,   {Oliva} E.,  2001b, \mn@doi [\aap]
  {10.1051/0004-6361:20000012}, \href
  {https://ui.adsabs.harvard.edu/abs/2001A&A...365...37M} {365, 37}

\bibitem[\protect\citeauthoryear{{Maloney}, {Hollenbach}  \&
  {Tielens}}{{Maloney} et~al.}{1996}]{1996Maloney}
{Maloney} P.~R.,  {Hollenbach} D.~J.,   {Tielens} A.~G.~G.~M.,  1996, \mn@doi
  [\apj] {10.1086/177532}, \href
  {https://ui.adsabs.harvard.edu/abs/1996ApJ...466..561M} {466, 561}

\bibitem[\protect\citeauthoryear{{Marinucci}, {Miniutti}, {Bianchi}, {Matt}  \&
  {Risaliti}}{{Marinucci} et~al.}{2013}]{2013Marinucci}
{Marinucci} A.,  {Miniutti} G.,  {Bianchi} S.,  {Matt} G.,   {Risaliti} G.,
  2013, \mn@doi [\mnras] {10.1093/mnras/stt1759}, \href
  {http://adsabs.harvard.edu/abs/2013MNRAS.436.2500M} {436, 2500}

\bibitem[\protect\citeauthoryear{{Martocchia} \& {Matt}}{{Martocchia} \&
  {Matt}}{1996}]{1996MartocchiaMatt}
{Martocchia} A.,  {Matt} G.,  1996, \mn@doi [\mnras] {10.1093/mnras/282.4.L53},
  \href {https://ui.adsabs.harvard.edu/abs/1996MNRAS.282L..53M} {282, L53}

\bibitem[\protect\citeauthoryear{{Matt}, {Perola}  \& {Piro}}{{Matt}
  et~al.}{1991}]{1991A&A...247...25M}
{Matt} G.,  {Perola} G.~C.,   {Piro} L.,  1991, \aap, \href
  {https://ui.adsabs.harvard.edu/abs/1991A&A...247...25M} {247, 25}

\bibitem[\protect\citeauthoryear{{Matt} et~al.,}{{Matt}
  et~al.}{1996}]{1996MNRAS.281L..69M}
{Matt} G.,  et~al., 1996, \mn@doi [\mnras] {10.1093/mnras/281.4.L69}, \href
  {https://ui.adsabs.harvard.edu/abs/1996MNRAS.281L..69M} {281, L69}

\bibitem[\protect\citeauthoryear{{Matt} et~al.,}{{Matt}
  et~al.}{1999}]{1999A&A...341L..39M}
{Matt} G.,  et~al., 1999, \aap, \href
  {https://ui.adsabs.harvard.edu/abs/1999A&A...341L..39M} {341, L39}

\bibitem[\protect\citeauthoryear{{Matt}, {Fabian}, {Guainazzi}, {Iwasawa},
  {Bassani}  \& {Malaguti}}{{Matt} et~al.}{2000}]{2000MNRAS.318..173M}
{Matt} G.,  {Fabian} A.~C.,  {Guainazzi} M.,  {Iwasawa} K.,  {Bassani} L.,
  {Malaguti} G.,  2000, \mn@doi [\mnras] {10.1046/j.1365-8711.2000.03721.x},
  \href {https://ui.adsabs.harvard.edu/abs/2000MNRAS.318..173M} {318, 173}

\bibitem[\protect\citeauthoryear{{McKaig}, {Ricci}, {Paltani}  \&
  {Satyapal}}{{McKaig} et~al.}{2021}]{McKaig:2021vw}
{McKaig} J.,  {Ricci} C.,  {Paltani} S.,   {Satyapal} S.,  2021, arXiv
  e-prints, \href {https://ui.adsabs.harvard.edu/abs/2021arXiv211100065M} {p.
  arXiv:2111.00065}

\bibitem[\protect\citeauthoryear{{Miniutti} \& {Fabian}}{{Miniutti} \&
  {Fabian}}{2004}]{2004MiniuttiFabian}
{Miniutti} G.,  {Fabian} A.~C.,  2004, \mn@doi [\mnras]
  {10.1111/j.1365-2966.2004.07611.x}, \href
  {https://ui.adsabs.harvard.edu/abs/2004MNRAS.349.1435M} {349, 1435}

\bibitem[\protect\citeauthoryear{{Murphy} \& {Yaqoob}}{{Murphy} \&
  {Yaqoob}}{2009}]{2009MNRAS.397.1549M}
{Murphy} K.~D.,  {Yaqoob} T.,  2009, \mn@doi [\mnras]
  {10.1111/j.1365-2966.2009.15025.x}, \href
  {http://adsabs.harvard.edu/abs/2009MNRAS.397.1549M} {397, 1549}

\bibitem[\protect\citeauthoryear{{Nandra} \& {Pounds}}{{Nandra} \&
  {Pounds}}{1994}]{Nandra:1994gu}
{Nandra} K.,  {Pounds} K.~A.,  1994, \mn@doi [\mnras]
  {10.1093/mnras/268.2.405}, \href
  {https://ui.adsabs.harvard.edu/abs/1994MNRAS.268..405N} {268, 405}

\bibitem[\protect\citeauthoryear{{Nandra}, {George}, {Mushotzky}, {Turner}  \&
  {Yaqoob}}{{Nandra} et~al.}{1997}]{Nandra:1997md}
{Nandra} K.,  {George} I.~M.,  {Mushotzky} R.~F.,  {Turner} T.~J.,   {Yaqoob}
  T.,  1997, \mn@doi [\apj] {10.1086/303721}, \href
  {https://ui.adsabs.harvard.edu/abs/1997ApJ...477..602N} {477, 602}

\bibitem[\protect\citeauthoryear{Netzer}{Netzer}{1990}]{Netzer1990}
Netzer H.,  1990, AGN Emission Lines.
Springer Berlin Heidelberg, Berlin, Heidelberg, pp 57--158,
  \mn@doi{10.1007/3-540-31625-6_2}, \url
  {https://doi.org/10.1007/3-540-31625-6_2}

\bibitem[\protect\citeauthoryear{{Netzer}}{{Netzer}}{2015}]{Netzer:2015rev}
{Netzer} H.,  2015, \mn@doi [\araa] {10.1146/annurev-astro-082214-122302},
  \href {http://adsabs.harvard.edu/abs/2015ARA%26A..53..365N} {53, 365}

\bibitem[\protect\citeauthoryear{{Nevalainen}, {David}  \&
  {Guainazzi}}{{Nevalainen} et~al.}{2010}]{2010Nevalainen}
{Nevalainen} J.,  {David} L.,   {Guainazzi} M.,  2010, \mn@doi [\aap]
  {10.1051/0004-6361/201015176}, \href
  {https://ui.adsabs.harvard.edu/abs/2010A&A...523A..22N} {523, A22}

\bibitem[\protect\citeauthoryear{{Odaka}, {Yoneda}, {Takahashi}  \&
  {Fabian}}{{Odaka} et~al.}{2016}]{2016MNRAS.462.2366O}
{Odaka} H.,  {Yoneda} H.,  {Takahashi} T.,   {Fabian} A.,  2016, \mn@doi
  [\mnras] {10.1093/mnras/stw1764}, \href
  {https://ui.adsabs.harvard.edu/abs/2016MNRAS.462.2366O} {462, 2366}

\bibitem[\protect\citeauthoryear{{Paltani} \& {Ricci}}{{Paltani} \&
  {Ricci}}{2017}]{2017A&A...607A..31P}
{Paltani} S.,  {Ricci} C.,  2017, \mn@doi [\aap] {10.1051/0004-6361/201629623},
  \href {http://adsabs.harvard.edu/abs/2017A%26A...607A..31P} {607, A31}

\bibitem[\protect\citeauthoryear{{Pancoast}, {Brewer}, {Treu}, {Park}, {Barth},
  {Bentz}  \& {Woo}}{{Pancoast} et~al.}{2014}]{Pancoast:2014uv}
{Pancoast} A.,  {Brewer} B.~J.,  {Treu} T.,  {Park} D.,  {Barth} A.~J.,
  {Bentz} M.~C.,   {Woo} J.-H.,  2014, \mn@doi [\mnras]
  {10.1093/mnras/stu1419}, \href
  {https://ui.adsabs.harvard.edu/abs/2014MNRAS.445.3073P} {445, 3073}

\bibitem[\protect\citeauthoryear{{Peterson} et~al.,}{{Peterson}
  et~al.}{2004}]{2004ApJ...613..682P}
{Peterson} B.~M.,  et~al., 2004, \mn@doi [\apj] {10.1086/423269}, \href
  {https://ui.adsabs.harvard.edu/abs/2004ApJ...613..682P} {613, 682}

\bibitem[\protect\citeauthoryear{{Ramos Almeida} \& {Ricci}}{{Ramos Almeida} \&
  {Ricci}}{2017}]{Ramos-Almeida:2017qq}
{Ramos Almeida} C.,  {Ricci} C.,  2017, \mn@doi [Nature Astronomy]
  {10.1038/s41550-017-0232-z}, \href
  {http://adsabs.harvard.edu/abs/2017NatAs...1..679R} {1, 679}

\bibitem[\protect\citeauthoryear{{Ricci}, {Ueda}, {Ichikawa}, {Paltani},
  {Boissay}, {Gandhi}, {Stalevski}  \& {Awaki}}{{Ricci}
  et~al.}{2014a}]{2014A&A...567A.142R}
{Ricci} C.,  {Ueda} Y.,  {Ichikawa} K.,  {Paltani} S.,  {Boissay} R.,  {Gandhi}
  P.,  {Stalevski} M.,   {Awaki} H.,  2014a, \mn@doi [\aap]
  {10.1051/0004-6361/201322701}, \href
  {https://ui.adsabs.harvard.edu/abs/2014A&A...567A.142R} {567, A142}

\bibitem[\protect\citeauthoryear{{Ricci}, {Tazaki}, {Ueda}, {Paltani},
  {Boissay}  \& {Terashima}}{{Ricci} et~al.}{2014b}]{Ricci:2014cf}
{Ricci} C.,  {Tazaki} F.,  {Ueda} Y.,  {Paltani} S.,  {Boissay} R.,
  {Terashima} Y.,  2014b, \mn@doi [\apj] {10.1088/0004-637X/795/2/147}, \href
  {https://ui.adsabs.harvard.edu/abs/2014ApJ...795..147R} {795, 147}

\bibitem[\protect\citeauthoryear{{Ricci}, {Ueda}, {Koss}, {Trakhtenbrot},
  {Bauer}  \& {Gandhi}}{{Ricci} et~al.}{2015}]{2015ApJ...815L..13R}
{Ricci} C.,  {Ueda} Y.,  {Koss} M.~J.,  {Trakhtenbrot} B.,  {Bauer} F.~E.,
  {Gandhi} P.,  2015, \mn@doi [\apjl] {10.1088/2041-8205/815/1/L13}, \href
  {http://adsabs.harvard.edu/abs/2015ApJ...815L..13R} {815, L13}

\bibitem[\protect\citeauthoryear{{Ricci} et~al.,}{{Ricci}
  et~al.}{2017a}]{2017ApJS..233...17R}
{Ricci} C.,  et~al., 2017a, \mn@doi [\apjs] {10.3847/1538-4365/aa96ad}, \href
  {https://ui.adsabs.harvard.edu/abs/2017ApJS..233...17R} {233, 17}

\bibitem[\protect\citeauthoryear{{Ricci} et~al.,}{{Ricci}
  et~al.}{2017b}]{Ricci2017ss}
{Ricci} C.,  et~al., 2017b, \mn@doi [\nat] {10.1038/nature23906}, \href
  {http://adsabs.harvard.edu/abs/2017Natur.549..488R} {549, 488}

\bibitem[\protect\citeauthoryear{{Ricci} et~al.,}{{Ricci}
  et~al.}{2018}]{Ricci:2018mp}
{Ricci} C.,  et~al., 2018, \mn@doi [Monthly Notices of the Royal Astronomical
  Society] {10.1093/mnras/sty1879}, \href
  {https://ui.adsabs.harvard.edu/abs/2018MNRAS.480.1819R} {480, 1819}

\bibitem[\protect\citeauthoryear{{Sambruna}, {Netzer}, {Kaspi}, {Brandt},
  {Chartas}, {Garmire}, {Nousek}  \& {Weaver}}{{Sambruna}
  et~al.}{2001}]{2001ApJ...546L..13S}
{Sambruna} R.~M.,  {Netzer} H.,  {Kaspi} S.,  {Brandt} W.~N.,  {Chartas} G.,
  {Garmire} G.~P.,  {Nousek} J.~A.,   {Weaver} K.~A.,  2001, \mn@doi [\apjl]
  {10.1086/318068}, \href {http://adsabs.harvard.edu/abs/2001ApJ...546L..13S}
  {546, L13}

\bibitem[\protect\citeauthoryear{{Shu}, {Yaqoob}  \& {Wang}}{{Shu}
  et~al.}{2010}]{2010ApJS..187..581S}
{Shu} X.~W.,  {Yaqoob} T.,   {Wang} J.~X.,  2010, \mn@doi [\apjs]
  {10.1088/0067-0049/187/2/581}, \href
  {https://ui.adsabs.harvard.edu/abs/2010ApJS..187..581S} {187, 581}

\bibitem[\protect\citeauthoryear{{Shu}, {Yaqoob}  \& {Wang}}{{Shu}
  et~al.}{2011}]{2011ApJ...738..147S}
{Shu} X.~W.,  {Yaqoob} T.,   {Wang} J.~X.,  2011, \mn@doi [\apj]
  {10.1088/0004-637X/738/2/147}, \href
  {https://ui.adsabs.harvard.edu/abs/2011ApJ...738..147S} {738, 147}

\bibitem[\protect\citeauthoryear{{Stalevski}, {Asmus}  \&
  {Tristram}}{{Stalevski} et~al.}{2017}]{2017MNRAS.472.3854S}
{Stalevski} M.,  {Asmus} D.,   {Tristram} K.~R.~W.,  2017, \mn@doi [\mnras]
  {10.1093/mnras/stx2227}, \href
  {http://adsabs.harvard.edu/abs/2017MNRAS.472.3854S} {472, 3854}

\bibitem[\protect\citeauthoryear{{Stalevski}, {Tristram}  \&
  {Asmus}}{{Stalevski} et~al.}{2019}]{2019MNRAS.484.3334S}
{Stalevski} M.,  {Tristram} K.~R.~W.,   {Asmus} D.,  2019, \mn@doi [\mnras]
  {10.1093/mnras/stz220}, \href
  {https://ui.adsabs.harvard.edu/abs/2019MNRAS.484.3334S} {484, 3334}

\bibitem[\protect\citeauthoryear{{Tanaka} et~al.,}{{Tanaka}
  et~al.}{1995}]{Tanaka:1995wm}
{Tanaka} Y.,  et~al., 1995, \mn@doi [\nat] {10.1038/375659a0}, \href
  {https://ui.adsabs.harvard.edu/abs/1995Natur.375..659T} {375, 659}

\bibitem[\protect\citeauthoryear{{Tanimoto}, {Ueda}, {Odaka}, {Kawaguchi},
  {Fukazawa}  \& {Kawamuro}}{{Tanimoto} et~al.}{2019}]{2019ApJ...877...95T}
{Tanimoto} A.,  {Ueda} Y.,  {Odaka} H.,  {Kawaguchi} T.,  {Fukazawa} Y.,
  {Kawamuro} T.,  2019, \mn@doi [\apj] {10.3847/1538-4357/ab1b20}, \href
  {https://ui.adsabs.harvard.edu/abs/2019ApJ...877...95T} {877, 95}

\bibitem[\protect\citeauthoryear{{Tanimoto}, {Ueda}, {Odaka}, {Ogawa},
  {Yamada}, {Kawaguchi}  \& {Ichikawa}}{{Tanimoto}
  et~al.}{2020}]{2020ApJ...897....2T}
{Tanimoto} A.,  {Ueda} Y.,  {Odaka} H.,  {Ogawa} S.,  {Yamada} S.,  {Kawaguchi}
  T.,   {Ichikawa} K.,  2020, \mn@doi [\apj] {10.3847/1538-4357/ab96bc}, \href
  {https://ui.adsabs.harvard.edu/abs/2020ApJ...897....2T} {897, 2}

\bibitem[\protect\citeauthoryear{{Tristram} et~al.,}{{Tristram}
  et~al.}{2007}]{2007A&A...474..837T}
{Tristram} K.~R.~W.,  et~al., 2007, \mn@doi [\aap]
  {10.1051/0004-6361:20078369}, \href
  {http://adsabs.harvard.edu/abs/2007A%26A...474..837T} {474, 837}

\bibitem[\protect\citeauthoryear{{Tristram}, {Burtscher}, {Jaffe},
  {Meisenheimer}, {H{\"o}nig}, {Kishimoto}, {Schartmann}  \&
  {Weigelt}}{{Tristram} et~al.}{2014}]{2014A&A...563A..82T}
{Tristram} K.~R.~W.,  {Burtscher} L.,  {Jaffe} W.,  {Meisenheimer} K.,
  {H{\"o}nig} S.~F.,  {Kishimoto} M.,  {Schartmann} M.,   {Weigelt} G.,  2014,
  \mn@doi [\aap] {10.1051/0004-6361/201322698}, \href
  {http://adsabs.harvard.edu/abs/2014A%26A...563A..82T} {563, A82}

\bibitem[\protect\citeauthoryear{{Ueda} et~al.,}{{Ueda}
  et~al.}{2007a}]{2007ApJ...664L..79U}
{Ueda} Y.,  et~al., 2007a, \mn@doi [\apjl] {10.1086/520576}, \href
  {https://ui.adsabs.harvard.edu/abs/2007ApJ...664L..79U} {664, L79}

\bibitem[\protect\citeauthoryear{{Ueda} et~al.,}{{Ueda}
  et~al.}{2007b}]{2007Ueda}
{Ueda} Y.,  et~al., 2007b, \mn@doi [\apjl] {10.1086/520576}, \href
  {https://ui.adsabs.harvard.edu/abs/2007ApJ...664L..79U} {664, L79}

\bibitem[\protect\citeauthoryear{{Uematsu}, {Ueda}, {Tanimoto}, {Kawamuro},
  {Setoguchi}, {Ogawa}, {Yamada}  \& {Odaka}}{{Uematsu}
  et~al.}{2021}]{2021Uematsu}
{Uematsu} R.,  {Ueda} Y.,  {Tanimoto} A.,  {Kawamuro} T.,  {Setoguchi} K.,
  {Ogawa} S.,  {Yamada} S.,   {Odaka} H.,  2021, arXiv e-prints, \href
  {https://ui.adsabs.harvard.edu/abs/2021arXiv210311224U} {p. arXiv:2103.11224}

\bibitem[\protect\citeauthoryear{{Venanzi}, {H{\"o}nig}  \&
  {Williamson}}{{Venanzi} et~al.}{2020}]{2020Venanzi}
{Venanzi} M.,  {H{\"o}nig} S.,   {Williamson} D.,  2020, \mn@doi [\apj]
  {10.3847/1538-4357/aba89f}, \href
  {https://ui.adsabs.harvard.edu/abs/2020ApJ...900..174V} {900, 174}

\bibitem[\protect\citeauthoryear{{Wada}, {Papadopoulos}  \& {Spaans}}{{Wada}
  et~al.}{2009}]{2009ApJ...702...63W}
{Wada} K.,  {Papadopoulos} P.~P.,   {Spaans} M.,  2009, \mn@doi [\apj]
  {10.1088/0004-637X/702/1/63}, \href
  {https://ui.adsabs.harvard.edu/abs/2009ApJ...702...63W} {702, 63}

\bibitem[\protect\citeauthoryear{{Walton}, {Nardini}, {Fabian}, {Gallo}  \&
  {Reis}}{{Walton} et~al.}{2013}]{Walton:2013qj}
{Walton} D.~J.,  {Nardini} E.,  {Fabian} A.~C.,  {Gallo} L.~C.,   {Reis} R.~C.,
   2013, \mn@doi [\mnras] {10.1093/mnras/sts227}, \href
  {https://ui.adsabs.harvard.edu/abs/2013MNRAS.428.2901W} {428, 2901}

\bibitem[\protect\citeauthoryear{{Walton} et~al.,}{{Walton}
  et~al.}{2019}]{Walton:2019gv}
{Walton} D.~J.,  et~al., 2019, \mn@doi [\mnras] {10.1093/mnras/stz115}, \href
  {https://ui.adsabs.harvard.edu/abs/2019MNRAS.484.2544W} {484, 2544}

\bibitem[\protect\citeauthoryear{{Weisskopf}, {Tananbaum}, {Van Speybroeck}  \&
  {O'Dell}}{{Weisskopf} et~al.}{2000a}]{2000Weisskopf}
{Weisskopf} M.~C.,  {Tananbaum} H.~D.,  {Van Speybroeck} L.~P.,   {O'Dell}
  S.~L.,  2000a, {Chandra X-ray Observatory (CXO): overview}.
pp 2--16, \mn@doi{10.1117/12.391545}

\bibitem[\protect\citeauthoryear{{Weisskopf}, {Tananbaum}, {Van Speybroeck}  \&
  {O'Dell}}{{Weisskopf} et~al.}{2000b}]{2000SPIE.4012....2W}
{Weisskopf} M.~C.,  {Tananbaum} H.~D.,  {Van Speybroeck} L.~P.,   {O'Dell}
  S.~L.,  2000b, {Chandra X-ray Observatory (CXO): overview}.
pp 2--16, \mn@doi{10.1117/12.391545}

\bibitem[\protect\citeauthoryear{{Wilkins} \& {Fabian}}{{Wilkins} \&
  {Fabian}}{2011}]{2011MNRAS.414.1269W}
{Wilkins} D.~R.,  {Fabian} A.~C.,  2011, \mn@doi [\mnras]
  {10.1111/j.1365-2966.2011.18458.x}, \href
  {http://adsabs.harvard.edu/abs/2011MNRAS.414.1269W} {414, 1269}

\bibitem[\protect\citeauthoryear{{XRISM Science Team}}{{XRISM Science
  Team}}{2020}]{2020arXiv200304962X}
{XRISM Science Team} 2020, arXiv e-prints, \href
  {https://ui.adsabs.harvard.edu/abs/2020arXiv200304962X} {p. arXiv:2003.04962}

\bibitem[\protect\citeauthoryear{{Yaqoob}}{{Yaqoob}}{2012}]{2012Yaqoob}
{Yaqoob} T.,  2012, \mn@doi [\mnras] {10.1111/j.1365-2966.2012.21129.x}, \href
  {https://ui.adsabs.harvard.edu/abs/2012MNRAS.423.3360Y} {423, 3360}

\makeatother
\end{thebibliography}






\appendix

\section{Emission lines}\label{ap:emission_lines}

\begin{figure}
\centering
\includegraphics[trim=0 0 0 0, clip,width=0.45\textwidth]{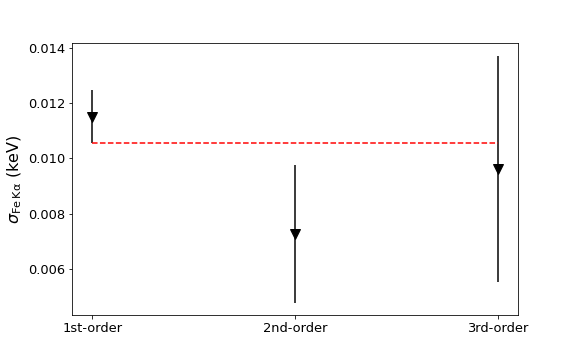}
\caption{Fe K$\alpha$ line widths against the spectral order of the {\it Chandra}/HEG spectrum. The aperture of the spectra is 3-pixel. The red dotted line marks the lower limit of the Fe K$\alpha$ line width of the 1-st order spectrum . Errors are plotted at a 90$\%$ confidence level.}\label{fig:width_ord}	
\end{figure}

Table\,\ref{t:lines} lists the emission lines identified by \citet{2001ApJ...546L..13S} and added in our fit, as our model does not reproduce them. 
Figure\,\ref{fig:width_ord} shows the variation of the Fe K$\alpha$ line widths with the spectral order of the {\it Chandra}/HEG spectrum.

\begin{table*}
\centering
 \begin{tabular}{l c c c c} 
 \hline
 \hline
\noalign{\smallskip}

\bf Line &\bf  Energy &{\bf Width} &{\bf Normalization} \\
\noalign{\smallskip}
 &  (keV) &(keV) & $(\rm 10^{-6} \times photons\: cm^{-2}\: s^{-1})$\\

 \hline
\noalign{\smallskip}
\noalign{\smallskip}
Ar XVIII & 3.106&$2.99 \times 10^{-3}$&$3.06$ \\
\noalign{\smallskip}
Ar XVIII &3.25&$ 4\times 10^{-3}$&1.2\\
\noalign{\smallskip}
Ca II-XIV + Ar XVIII & 3.69&$ 5\times 10^{-2}$&$2.87$\\
\noalign{\smallskip}
Ar XVIII & 3.9&$1 \times 10^{-3}$&$0.94$ \\
\noalign{\smallskip}
Ca II & 5.41&$1.03 \times 10^{-4}$&$3.51$\\

\hline
\hline
 \end{tabular}
 \caption{Emission lines identified by \citet{2001ApJ...546L..13S} and added in our model to fit the X-ray spectrum of Circinus. For details about the emission lines detected on the contamination sources, see \citet{2014ApJ7...91...81A}. } \label{t:lines}
\end{table*}

\section{Contamination spectrum} \label{ap:contamination}

Here, we report the best-fitting parameters of the off-nuclear Compton-scattered continuum, modeled with \textsc{mytorus}, and the AGN scattered emission. The photon-indexes were tied to the one of the primary emission, and we allow to vary the column density of the reflector $N_{\rm H,cont}$, the inclination angle, and the normalizations. Our results are consistent with those obtained by \citet{2014ApJ7...91...81A}.

\begin{table*}
\centering

 \begin{tabular}{lcc} 
 \hline
 \hline
\noalign{\smallskip}

\bf Component &\bf  Parameters  & \\
\noalign{\smallskip}
 \hline
 \noalign{\smallskip}
\noalign{\smallskip}

 \textsc{mytorus}& $ N_{\rm H,cont}$ ($\rm cm^{-2}$)& $(9.8^{+nc}_{-6})  \times 10^{24}$ \\
 \noalign{\smallskip}
&Inclination angle (deg)&  $78.7^{+4.3}_{-3.7}$\\
 \noalign{\smallskip}
 &$N_{\rm s,cont}$ ($\rm ph\: keV^{-2}\: s^{-1}\: cm^{-2}$)& $0.066^{+0.007}_{-0.002}$ \\
 \noalign{\smallskip}
Scattered pl& $N_{\rm cont}$ ($\rm ph\: keV^{-2}\: s^{-1}\: cm^{-2}$) & $(1.4^{+0.6}_{-0.3})\times 10^{-4}$\\

 \noalign{\smallskip}
\hline
 \noalign{\smallskip}
 \end{tabular}
 \caption{Parameters of the best-fitting models obtained for contamination model. $N_{\rm H,cont}$ is the equatorial column density, $N_{\rm s,cont}$ is the normalization of the  \textsc{mytorus} component and $N_{\rm cont}$ is the normalization of the scattered powerlaw.  $ nc$ means not constrained.} \label{t:cont_fits}
\end{table*} 

\section{Broadband fit without the contamination spectrum} \label{ap:fit_nocont}

\begin{figure} 
\centering
 \includegraphics[trim=0 160 0 40, clip,width=0.48\textwidth]{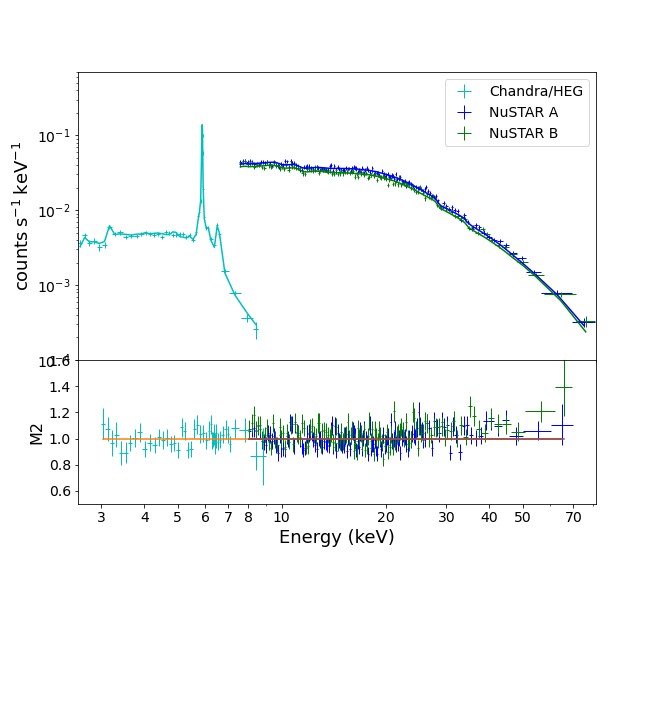} 

 \caption[]{\textit{NuSTAR} FPMA and FPMB spectra (blue and green, respectively) and \textit{Chandra}/HEG spectrum (cyan), altogether with the best fitting model \textsc{M1}  + \textsc{G}$_{\rm Fe}$ (i.e., without considering the contamination affecting the \textit{NuSTAR} spectra). The bottom panel shows the ratio between the data and \textsc{model\_2}. \label{fig:overall_fit_nocont} }  
 \end{figure}

Here, we report the best-fitting parameters (see Table\,\ref{t:nustar_nocont})of the joint fit between \textit{Chandra}/HEG and {\it NuSTAR} to the model \textsc{M1}  + \textsc{G}$_{\rm Fe}$ (without considering the contamination affecting the {\it NuSTAR} spectrum. See Figure\,\ref{fig:overall_fit_nocont}).

\begin{table*}
\centering
 \begin{tabular}{lcc} 
 \hline
 \hline
\noalign{\smallskip}

\bf Component &\bf  Parameters &{\bf M1 + \textsc{G}$_{\rm Fe}$} \\
\noalign{\smallskip}
 \hline
 \noalign{\smallskip}
\noalign{\smallskip}
 Cone model&$\Gamma$&$1.6^{+0.004}_{- nc}$ \\
 \noalign{\smallskip}
&$  N_{\rm H,d}$ ($\rm cm^{-2}$)&$\rm 6.1^{+0.72}_{-0.6}\times 10^{24} $ \\
 \noalign{\smallskip}
&$ N_{ \rm H,c}$ ($\rm cm^{-2}$)&$\rm 0.84^{+0.15}_{-0.067}\times 10^{22} $\\
 \noalign{\smallskip}
&$CF$&$0.54 \pm 0.01$ \\
 \noalign{\smallskip}
&$N_l$ ($\rm ph\: keV^{-2}\: s^{-1}\: cm^{-2}$) & $0.21^{+0.02}_{-0.01}$\\
 \noalign{\smallskip}
 &$N_s$ ($\rm ph\: keV^{-2}\: s^{-1}\: cm^{-2}$)&$0.19 \pm 0.01 $ \\
 \noalign{\smallskip}
Scattered pl& $N/N_c$ &$2.1 \pm 0.26\times 10^{-3}$ \\

\noalign{\smallskip}
\hline
\noalign{\smallskip}
$\rm stats/dof$ && 1951/2410  \\
 \noalign{\smallskip}
\hline
 \noalign{\smallskip}
 \end{tabular}
 \caption{Parameters of the best-fitting models obtained for the \textit{Chandra}/HEG and {\it NuSTAR}, without considering the contamination spectrum. $\Gamma$ is the photon index, $ N_{\rm H,d}$ is the equatorial column density of the flared disk,  $ N_{\rm H,c}$ is the radial column density of the cone shell, and $CF$ is the covering factor of the flared disk. $N_l$ and $N_s$ are the normalizations of the emission lines  and the scattered component of the AGN, respectively. $N/N_c$ is the ratio between normalizations of the scattered powerlaw and the continuum. The normalization of the continuum is tied to the normalization of the scattered emission ($N_c=N_s$). $ nc$ means not constrained.  } \label{t:nustar_nocont}
\end{table*}

\bsp	
\label{lastpage}
\end{document}